\documentclass[twocolumn,times]{aastex631} 
\usepackage{amsmath}
\usepackage{mathtools}
\usepackage{eucal}
\usepackage{amssymb}
\usepackage{enumitem}
\usepackage{svg}
\usepackage{comment}
\received{\today}

\newcommand{\p}{\partial} 
\newcommand{\SMbh}{M_\bullet}

\newcommand{\mstar}{m_\star}

\newcommand{\Msun}{M_{\odot}}
\newcommand{\TL}{\Tbb^{L}}

\newcommand{\E}{\mathcal{E}}

\newcommand{\Tgw}{T_{\rm gw}}

\newcommand{\beq}{\begin{equation}} 
\newcommand{\eeq}{\end{equation}} 
\newcommand{\rmd}{ {\rm d}  }

\newcommand{\Tbb}{T_{\rm 2b}}
 
\newcommand{\Tkep}{T_{\rm Kep}}

\newcommand{\SMBhMW}{M_{\bullet {\rm mw}}}

\newcommand{\lnL}{\ln{\Lambda}}

\newcommand{\xini}{\mbox{$x_{\rm ini}$}}

\newcommand{\scF}{\mathcal{F}}
\newcommand{\scN}{\mathcal{N}}

\newcommand{\Nf}{N_{\!f}}
\newcommand{\Nfo}{N_{\!f0}} 
\newcommand{\gamf}{\gamma_{\!f}}
\newcommand{\mf}{m_{\!f}}

\newcommand{\scNgw}{\scN_{\rm gw}}
\newcommand{\scNi}{\scN_i} 
\newcommand{\rpf}{r_{p {\rm f}}}
\newcommand{\rtid}{r_{\rm tid}} 
\newcommand{\Remri}{\Gamma}

\newcommand{\rcap}{r_{\rm cap}} 

\newcommand{\rpgw}{r_{p, {\rm gw}}}
\newcommand{\rgw}{r_{\rm gw}} 
\newcommand{\ggw}{g_{\rm gw}}
\newcommand{\Ei}{{\rm Ei}} 
\newcommand{\gamE}{\gamma_{\rm E}}
\newcommand{\xfin}{x_{\rm fin}}

\graphicspath{{graphs/}}

\begin{document}

\title{Semi-Analytical Fokker Planck Models for Nuclear Star Clusters}

\correspondingauthor{Karamveer Kaur}
\email{karamveerkaur30@gmail.com}

\author[0000-0001-6474-4402]{Karamveer Kaur}
\affiliation{Technion - Israel Institute of Technology, Haifa, 3200002, Israel}
\affiliation{Racah Institute of Physics, The Hebrew University of Jerusalem, 9190401, Israel}

\author[0000-0002-7420-3578]{Barak Rom}
\affiliation{Racah Institute of Physics, The Hebrew University of Jerusalem, 9190401, Israel}

\author[0000-0002-1084-3656]{Re'em Sari}
\affiliation{Racah Institute of Physics, The Hebrew University of Jerusalem, 9190401, Israel}

\shorttitle{Fokker Planck Models for Nuclear Star Clusters}
\shortauthors{Kaur et al.}

\begin{abstract}

We study the dynamics of nuclear star clusters, the dense stellar environments surrounding massive black holes in the centers of galaxies.  
We consider angular momentum diffusion due to two-body scatterings among stellar objects and energy advection due to gravitational wave emission upon interaction with the central massive black hole. Such dynamics is described by a two-dimensional Fokker-Planck equation in energy-angular momentum space.
Focusing on the transition between the diffusion-dominated region and the advection-dominated one, we utilize self-similarity to obtain a full solution for the Fokker-Planck equation.
This solution provides the density and flux of the stellar objects in nuclear star clusters. This improves the rate estimates for extreme mass-ratio inspirals, and has interesting implications for a new class of galactic center transients called quasi-periodic eruptions. 
\end{abstract}



\section{Introduction} 

Centers of many galaxies harbor massive black holes (MBHs) surrounded by dense nuclear star clusters (NSCs) \citep{Kormendy_Ho_2013,Neumayer_2020}. Different dynamical processes can render stars and compact objects on orbits leading to the close proximity of MBH, resulting in a rich variety of electromagnetic and gravitational wave (GW) transients \citep{Tal_2017}. 
Extreme-mass ratio inspirals (EMRIs) of  stellar-mass objects, especially stellar black holes (BHs), are one of the promising sources for the upcoming low-frequency GW detectors - Laser Interferometer Space Antenna (LISA) and TianQin \citep{Amaro-Seoane_2012lisa,eLISA_Consort2013,Amaro-Seoane_2017,Mei_2021}. EMRIs are expected to spend upto $\sim 10^5$ orbital cycles while shining in the observable wavebands ($\sim 10^{-4}-1$~Hz) before making a final plunge into the central MBH \citep{Peters_1964}. This makes them an excellent probe of space-time geometry around MBH to test general relativity and alternative theories of gravity in the strong field regime \citep{Ryan_1995,Ryan1997,Glampedakis_2006,Barack_2007,Gair_2013}. Predicted occurrence rates of EMRIs are sensitive to inherent characteristics of NSCs and hence can unveil the underlying MBH demographics, structure of NSCs, and dynamical processes operating therein responsible for the formation of these sources \citep{Amaro-Seoane_2007,Gair_2010,Gair_Sesana_2011}.   

Apart from the initial focus on compact object EMRIs, main sequence (MS) EMRIs for GW-induced inspiral of stars have been receiving attention recently \citep{Linial_Sari_2017,Metzger_2022,Linial_Sari_2023}. The MS EMRIs (or their progenitors) interacting with an accretion disk around central MBH \citep{Xian_2021,Linial_Metzger_2023,Linial_2024_uv,Tagawa_2023} are possibly associated with a new class of short period (few hours) soft X-ray transients called quasi-periodic eruptions (QPEs) \citep{Miniutti_2019,Giustini_2020,Arcodia_2021,Arcodia_2024_newQPEs}, though there are other possible formation scenarios of these transients \citep{Raj2021,Pan_2022_viscosity_model,Kaur_Stone_2023}. Even, BH EMRIs (or their progenitors) $+$ disk interactions have been associated with QPEs \citep{Franchini_2023}. Theoretical investigations of rates of these transients are motivated by the recent advances in X-ray astronomy, with the current and future survey telescopes, like eROSITA \citep{Predehl_2021}, wide-field X-ray telescope onboard Einstein Probe \citep{Yuan_2022}, Wide Field Imager onboard Athena \citep{Nandra_2013} (and may be even UV missions \citep{Linial_2024_uv}, like ULTRASAT \citep{Sagiv_2014}, UVEX \citep{Kulkarni_2021}).

Different dynamical channels for EMRI formation have been proposed in the past. In the classical scenario driven by angular momentum  relaxation, mutual two-body scatterings channel objects onto high eccentricity orbits so that there is efficient energy dissipation by GW emission near periapses \citep{Hils_1995,Sigurdsson_1997,Freitag_2001,Hopman_Alexander_2005}. Tidal disruption of binaries owing to Hills mechanism \citep{Hills1988,Yu_Tremaine_2003}, deposits objects onto tightly bound high eccentricity orbits, that can eventually become EMRIs \citep{Miller_2005,Fragione_Sari_2018,Raveh_2021}. Other collisionless effects like Lidov-Kozai mechanism due to a central MBH binary \citep{Bode_2014,Mazzolari_2022,Naoz_2022}, can be important as well. Dissipative interactions with an accretion disk can lead to capture and migration in the disk, eventually funneling a population of low-eccentricity EMRIs \citep{Levin_2007,Pan2021}. Further, gravitational instabilities for stellar disks may excite stars onto highly eccentric orbits \citep{Madigan_2009,Kaur_2018}, leading to higher rates of nuclear transients. Massive perturbers like giant molecular clouds can also play an important role in channeling objects onto EMRI orbits \citep{Perets2007}.

In this work, we focus on the classical channel of scattering-driven angular momentum relaxation, which itself has many uncertainties. Dynamical uncertainties include plunge-EMRI dichotomy which is more pronounced for \emph{less-extreme} mass ratio inspirals around central intermediate-mass black holes (IMBHs) \citep{Qunbar_Stone_2023}, and the nature of segregation regime - strong or weak - favored by real NSCs \citep{Hopman_2006,Keshet_2009,Aharon_Perets_2016,Linial_Sari_22}. Further, various poorly-constrained factors like mass function of stellar BHs in NSCs, timescales of replenishment of sources in the nuclei of low-mass galaxies, and the low-mass end of MBH mass function, further add to uncertainties in EMRI rates and characteristics \citep{Babak2017,Gair_2017,Broggi_2022}.

Numerous incoherent two-body (2B) scatterings among stellar objects provide fluctuations over the smooth background gravitational  potential, and drive relaxation of stellar energies 
and angular momenta over 2B timescales $\Tbb$ (for near-circular orbits), much longer than the dynamical timescale $\Tkep$. However, highly eccentric orbits that channel most EMRIs, undergo a much faster diffusion in angular momentum (and hence orbital periapse $r_p$) over times $\TL \simeq \Tbb 2 r_p/r$ \citep{Binney_Tremaine,Merritt_2013}, where $r$ represents the length of semi-major axis. Eventually as orbital eccentricities increase, energy dissipation due to GW losses (that occur on timescales $\Tgw$) near periapsidal passage dominates over the impact of 2B scatterings. In the inner-most regions of NSC lying deep within the radius of influence $r_h$ of MBH \citep{Hopman_Alexander_2005}, sources can enter a GW dominated regime ($\Tgw < \TL$) which leads to gradual shrinkage of semi-major axis $r$, while roughly conserving $r_p$ \citep{Peters_1964}. This would ultimately lead to an \emph{observable} EMRI as the emitted GW frequency enters LISA waveband for an MBH of mass $\SMbh \simeq 10^{4-7} \Msun$.

There is a vast literature on numerical studies of the relaxation-driven formation of EMRIs (eg., \citealt{Hopman_Alexander_2005, Amaro_Seoane_Preto_2011,Merritt_2015,Bar-Or_2016,Amaro-Seoane_2018,Broggi_2022}). Most of these works numerically solve a two-dimensional (2D) Fokker-Planck (FP) equation in energy-angular momentum space considering diffusion due to 2B scatterings and advection due to GW emission. In the current study, we develop a novel semi-analytical approach to solve the steady-state 2D Fokker-Planck equation, and determine the orbital distribution of EMRI progenitors, which finally decides the rates and characteristics of observable EMRIs. The method relies on the self-similarity of the problem for a star cluster with a power-law density profile. This greatly simplifies the 2D problem of EMRI formation to an effective one-dimensional (1D) scale-free ordinary differential equation, that is solved numerically. The significance of our approach lies in the ability to express density distribution and flux of EMRI progenitors in terms of a simplified analytical model. This makes our semi-analytical model extremely handy for exploring the multi-dimensional parameter space associated with astrophysical NSCs.

We describe the physical set-up of the model NSC and 2D Fokker-Planck formalism associated with relevant physical mechanisms in \S~\ref{sec_FP_formal}. The simplified analytical model for the orbital distribution of EMRI progenitors is described in  \S~\ref{sec_ana_model}. Then, we describe the more general approach of our semi-analytical model in \S~\ref{sec_SA_model}, which includes reduction to the final 1D problem. We present the results based on this approach in \S~\ref{sec_res}. We finally conclude in \S~\ref{sec_discus}, while discussing implications for the rates of EMRIs and QPEs.

\section{Physical setting and Fokker-Planck formalism} 

\label{sec_FP_formal} 

We consider a two-population model of an NSC with a central MBH of mass $\SMbh$. The NSC hosts two types of stellar objects -- stars and stellar-mass BHs.  To describe the dynamics of EMRI formation in a general scenario, we refer to the dominant scatterers as \emph{field stars} of mass $\mf$ and the population of interest for EMRI formation as \emph{subject stars} of mass $m$. Thermally relaxed star clusters are expected to have a more compact distribution for heavier BHs due to mass segregation \citep{Bahcall_Wolf_1977}. Unless stars are highly dominant, it is feasible that inner regions have BHs as dominant scatterers; while in the outer regions, scattering due to stars is expected to dominate \citep{Alexander_Hopman2009,Linial_Sari_22}. Hence there are four possible combinations of $\{$field, subject$\}$ stars as both MS and BH EMRIs are the subjects of interest.

Since two-body scatterings alter stellar orbital energies $\E = G \SMbh/(2 r)$ (and hence, semi-major axes $r$) over longer 2B-relaxation time $\Tbb$ \citep{Binney_Tremaine} \footnote{In this convention, we call specific energy $\E$ as \emph{energy} which is positive for bound orbits.}, formation of EMRIs occurs as stellar orbits relax in angular momenta $L$ and a fraction of them are scattered onto high eccentricity orbits capable of emitting GWs efficiently near their periapses $r_p$ \citep{Hopman_Alexander_2005}. The relaxation in angular momenta $L = \sqrt{2 G \SMbh r_p}$ (and hence $r_p$) occurs over a much shorter timescale $\TL = (2 r_p/r) \Tbb $. Hence, the energy relaxation is ignored and distributions of both field and subject stars are assumed to be fixed in $r$. We choose the following form for 2B timescale \citep{Merritt_2011,Merritt_2013,Bortolas2019}: 
\beq 
\Tbb(r) = \frac{3\sqrt{2} \pi^2}{32 C}  \frac{\Tkep(r)}{\lnL} \bigg( \frac{ \SMbh}{\mf} \bigg)^2 \frac{1}{\Nf(r)} \,
\label{Tbb}
\eeq 
where dynamical timescale $\Tkep = \sqrt{r^3/(G \SMbh)}$ and $\lnL$ is the Coulomb logarithm.  We choose $\Nf(r) = \Nfo (r/r_h)^{3-\gamf}$, the number of field particles with semi-major axis $<r$, where $r_h$ is the radius of influence of MBH \footnote{Later in this section, we also designate a similar number profile $N(r) = N_0 (r/r_h)^{3 - \gamma}$ for the subject stars. Strictly speaking, the implied  power-law density profile with index $\gamma$ for subject stars is valid only at higher angular momentum or $r_p$, where both angular momentum relaxation and GW losses are negligible.}. The numerical factor $C$ is a function of $\gamf$ with $C \in [0.98,1.61]$ for $\gamf \in [1.5, 3]$; see appendix~A of \citet{Bortolas2019}. We use $C=1.35$ which is appropriate for the Bahcall-Wolf (BW) profile of field stars ($\gamf = 7/4$; \citealt{Bahcall_1976}) and ignore its weak variation within the physically interesting range of $\gamf$. Similar to earlier studies, we consider frozen-in approximation with a thermal distribution in $L$ of the background field stars \citep{Merritt_2013}. Here we neglect the collective effects induced by resonant relaxation \citep{Rauch_Tremaine_1996}, as they do not impact the overall EMRI rates significantly \citep{Alexander_2017RR}.

 As subject stars are scattered onto high eccentricity orbits, orbital energy losses due to GW emission near $r_p$ can become important  and occur over GW timescale \citep{Peters_1964}:
 \beq 
\Tgw(r,r_p) \equiv {r \over |\dot r|} = \frac{96 \sqrt{2}  }{85} \frac{R_s}{c} \frac{\SMbh}{m} \bigg( \frac{r_p}{R_s} \bigg)^4 \sqrt{\frac{r}{r_p}}\,
\label{Tgw}
 \eeq 
  where $R_s= 2 G \SMbh/c^2$ is the Schwarzschild radius. For sufficiently small $r_p$, $\Tgw < \TL$ so that GW-driven orbital shrinkage in $r$ begins to dominate the scattering-driven diffusion in $r_p$. Here we ignore the angular momentum losses due to GWs, which become important only later on as eccentricities decay. For ease of notation, we will denote the GW timescale $\Tgw(r) \equiv \Tgw(r,r) $ for the case of a circular orbit ($r_p = r$).

So the relevant dynamics of subject stars includes relaxation in angular momentum $L$ driven by two-body scatterings, and energy dissipation due to GW emission. The Fokker Planck (FP) equation for the evolution of subject star 2D density $\scN(r,r_p)$ in $\{r,r_p \}$-plane, accounts for both these effects and is given as: 
\beq 
\frac{\p \scN}{\p t} =  \frac{\p \scF_p}{\p r_p} + \frac{\p \scF_r}{\p r} \,.  
\label{FP_r_rp}
\eeq 
Here the flux stream densities along $r_p$ and $r$ respectively are $\scF_p$ and $\scF_r$: 
\beq 
\begin{split}
&\scF_p = \frac{r}{2 \, \Tbb(r)}  \, r_p \frac{\p \scN}{\p r_p}  = \frac{r_h^{\gamf - 3/2}}{2 \, \Tbb(r_h)} \, r^{5/2 - \gamf} \, r_p \frac{\p \scN}{\p r_p}  \\[1ex]
& \scF_r = \frac{r \scN}{\Tgw(r,r_p)}  = \frac{r_h^4}{\Tgw(r_h)} \frac{\sqrt{r}}{ r_p^{7/2} } \scN 
\end{split} 
\label{fluxes_Fp_Fr}
\eeq 
The first term on the right side of the partial differential equation~(\ref{FP_r_rp}) physically corresponds to angular momentum relaxation, and combines both drift and diffusion terms by employing the fluctuation-dissipation theorem \citep{Lightman_Shapiro1977,Merritt_2013} \footnote{Our notation for relaxation in $r_p$ relates directly to \citealt{Merritt_2013} (equation~6.29) with $ \mathcal{R} = L^2/ L_c^2 = 2 r_p/r$ and $\Tbb = L_c^2/(2 \langle (\Delta L)^2  \rangle) = \mathcal{D}^{-1}$. Here $L_c = \sqrt{G \SMbh r}$ is the angular momentum of a circular orbit, and $\langle (\Delta L)^2  \rangle$ is the orbit-averaged diffusion coefficient for $L$.}. The second term accounts for the energy dissipated by GW emission.

As is evident from the FP equation~(\ref{FP_r_rp}), the first term for $r_p$ diffusion dominates for large $r_p$, while the second/advective term in $r$ dominates for small $r_p \lesssim \rpgw$. Here $\rpgw$ is an approximate threshold in $r_p$ below which energy loss by GW emission dominates the orbital evolution over two-body scatterings. We call this boundary $r_p = \rpgw$ separating the two dynamical regimes as \emph{GW loss cone boundary}, and its explicit form can be evaluated by equating the $L$-relaxation time $\TL$ and GW emission time $\Tgw$. This suggests the condition $\Tbb(r) 2 r_p/r = \Tgw(r,r_p)$ for $r_p = \rpgw$, which can also be obtained by naive dimensional analysis of the equation~(\ref{FP_r_rp}). Using equations~(\ref{Tbb}) and (\ref{Tgw}) in this condition, gives an explicit form of this boundary: 
\beq 
\begin{split} 
\rpgw(r) &= r_h \bigg( \frac{2 \Tbb(r_h)  }{ \Tgw(r_h)  }  \bigg)^{2/5} \bigg( \frac{r_h}{r} \bigg)^{2(3 - \gamf)/5}  \\[1ex]
  &= R_s \bigg[   \frac{\xi}{\lnL} \frac{m}{ \SMbh} \bigg( \frac{\SMbh}{\mf} \bigg)^{2} \frac{1}{\Nfo} \bigg]^{2/5} 
           \bigg( \frac{r_h}{r}\bigg)^{2(3 - \gamf)/5}  \\[1ex] 
 \mbox{with the} & \mbox{ numerical factor  } \xi = \frac{85 \pi^2}{ 256 \sqrt{2} C } \simeq 1.7    \\       
\end{split} 
\label{rp0}
 \eeq

\begin{figure}
    \includegraphics[width=0.55\textwidth,trim={4cm 0 3cm 0}]{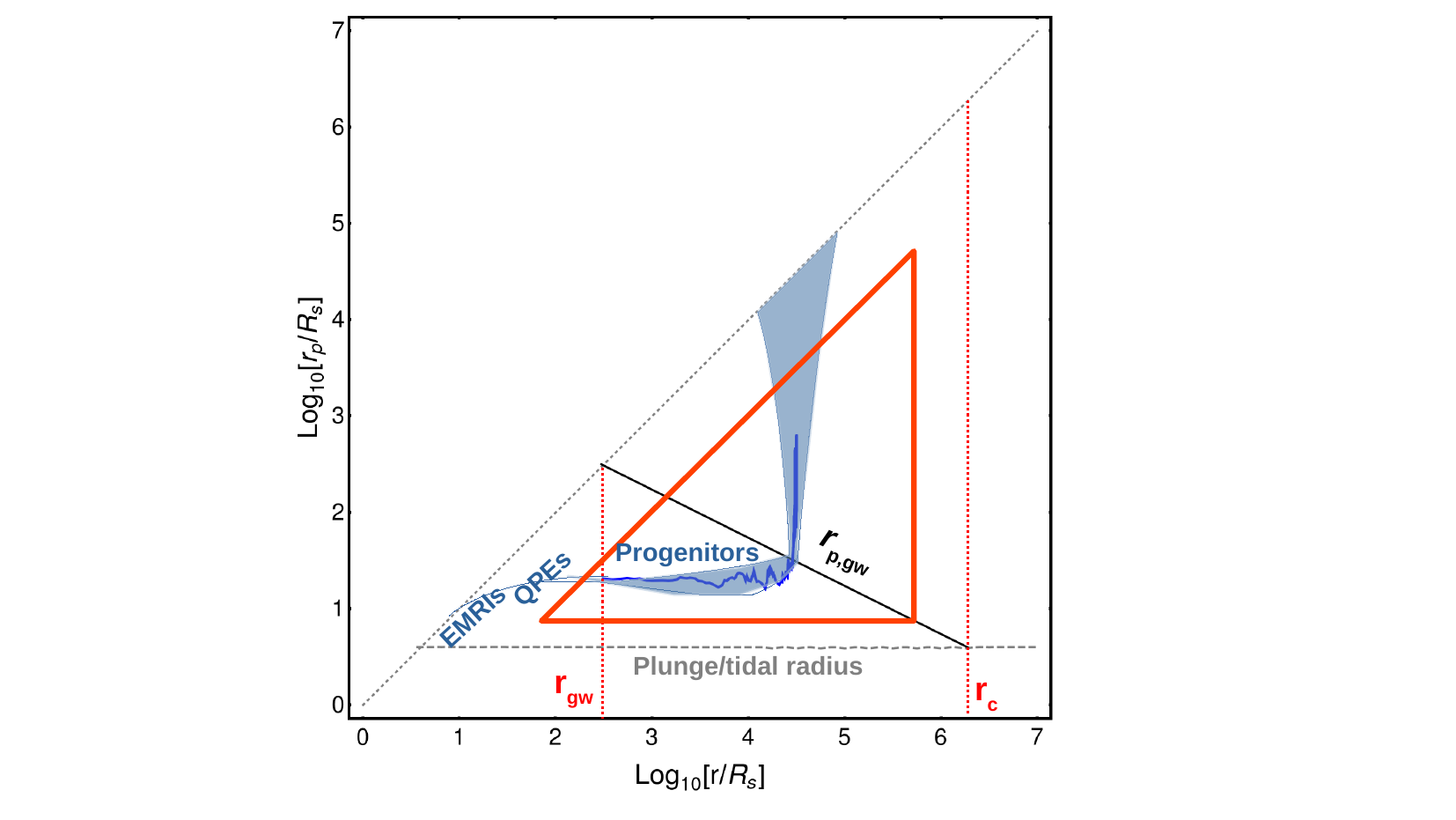}
    \caption{A schematic of the evolution of a typical orbit (in light blue) resulting in an EMRI, is shown. The loss cone boundary $\rpgw$ (solid black line) separates the scattering dominated regime above it, from GW dominated regime below. Our calculation is valid inside the region enclosed by the solid red boundary line. We schematically show these boundaries --
    $r \ll r_c$, $r_p \ll r $. The lower limit on $r_p$ is $\rcap $ which is either the tidal radius $\rtid$ (for stars) or $ 4 R_s$ for BHs. Here we choose stars as field particles of mass $m_f = 1 \Msun$ following a BW density profile around a central MBH of mass $\SMbh = 10^6 \Msun$. Location of EMRIs corresponds to the entry of the source into LISA waveband for $r \lesssim 20 R_s$, while that of QPEs corresponds to $r \lesssim 100 R_s$ so that orbital period is less than a day. 
      }
      \label{fig_schematic}
\end{figure}

The loss cone boundary $\rpgw$ separates the GW-dominated regime or GW loss cone ($r_p \leq \rpgw$) from diffusion-dominated regime ($r_p > \rpgw$). For BW profile of field stars ($\gamf = 7/4$), $\rpgw \propto 1/\sqrt{r}$ \citep{Sari_Fragione_2019}.  We highlight this boundary $r_p = \rpgw$ (in black) that separates the two dynamical regimes in figure~\ref{fig_schematic}. The figure shows a schematic of evolution of a typical stellar orbit, that evolves into an EMRI, in $\{r,r_p\}$-plane. Further, for highly bound orbits with $r < \rgw$ (that satisfies the condition $\rpgw(\rgw) = \rgw$), GW emission begins to dominate for even circular orbits. 

\emph{Region of validity of our approach}: For high orbital energies, there exists a critical $r = r_c$ for which $\rpgw(r_c)$ equals capture radius $\simeq4 R_s$ for BHs (or tidal disruption radius $\rtid$ for stars). This critical semi-major axis $r = r_c$ is shown as a dotted red vertical line (on the right) in the figure~\ref{fig_schematic}. Earlier studies also suggest that most EMRIs are sourced by subject stars with $r \leq r_c$ \citep{Hopman_Alexander_2005,Raveh_2021,Sari_Fragione_2019}, while \{BHs, stars\} at higher $r$ lead to \{plunges, tidal disruption events (TDEs)\}, when scattered onto high eccentricity orbits. This simplified picture is not complete and for a central IMBH ($\SMbh \lesssim 10^4 \Msun$) there can be a finite contribution for BH EMRIs coming from very large $r \gg r_c$ \citep{Qunbar_Stone_2023}. However, in the current study, we neglect the plunges/TDEs (and associated loss cones) and study the dynamics of EMRI formation around a central MBH for low-energy orbits with $r \ll r_c$. 

Non-consideration of plunges/TDEs further constrains the region of validity of our approach, such that $r_p$ is well above the plunges/TDEs loss cone marked by capture radius $\rcap$ (dashed gray horizontal line). For stars (of mass $\mstar$ and size $R_{\star}$), capture radius $\rcap = \rtid = R_{\star} (\SMbh/\mstar)^{1/3}$ is the tidal disruption radius as the stars with $r_p < \rtid$ become TDEs. For BHs, the usual choice of the capture radius $\rcap \simeq 4 R_s$ \footnote{This corresponds to a critical stellar angular momentum $4 G \SMbh/c$ for which the peak of effective potential in Schwarzschild geometry becomes zero. }. 

Also, our FP treatment of diffusion in $r_p$ (by combining the drift and diffusion terms as explained earlier below equation~\ref{fluxes_Fp_Fr}) assumes high eccentricities $r_p \ll r$. The region in $\{r,r_p \}$-plane which satisfies the above three constraints is shown qualitatively as the area enclosed by the solid red boundary in figure~\ref{fig_schematic}.  

\emph{Boundary conditions}: We intend to solve for a steady-state solution $\scN(r,r_p)$ of the FP equation~(\ref{FP_r_rp}) with a suitable boundary condition at the high angular momentum end $L \simeq L_c$ or equivalently $r_p \simeq r/2$  \footnote{Note that our simplified units suitable for eccentric orbits $L = \sqrt{2 G \SMbh r_p}$ yield a coordinate boundary $r_p = r/2$, corresponding to circular orbits. This choice of limiting boundary gives accurate form of 2D number density $\scN$ and flux density $\scF_p$ in diffusive regime, as given by equations~(\ref{scNi}) and (\ref{Fp_approx}) respectively. } (in the diffusive regime $r_p \gg \rpgw$), which remains relatively \emph{undisturbed} from both two-body scatterings and GW emission. We further demand this asymptotic solution $\scNi(r,r_p)$ to be consistent with a power-law profile $N(r) = N_0 (r/r_h)^{3 - \gamma}$ for the subject stars, which sets its $r$-dependence: $\scNi(r,r/2) = 2 N(r)/r^2$. Its $r_p$-dependence can be deduced explicitly by ignoring GW emission (by setting $\scF_r = 0$ in equation~\ref{FP_r_rp}), which reduces the steady-state equation to $\p \scF_p/\p r_p = 0$. This leads generally to a logarithmic profile for a non-zero flux $\scF_p$ \footnote{We will see later that it is not possible to have a non-zero flux $\scF_p$  for subject stars with shallow density profiles $\gamma \leq 3/2$. For a steady-state solution with $\gamma = 3/2$, one needs to choose a zero flux condition $\scF_p \rightarrow 0$ as $r_p \rightarrow r$, which makes the asymptotic solution $\scNi$ independent of $r_p$.} that gives \citep{Cohn_Kulsrud_1978}: 
\beq 
\scNi(r,r_p) = \displaystyle{\frac{2 N_0}{ r_h^{3-\gamma}} \, r^{1-\gamma} \, \frac{ \ln{(r_p/\rpgw)} + c_0}{ \lnL_0 + c_0   } } \; ,   \; \,  \mbox{for} \; r_p \gg \rpgw
\label{scNi}
\eeq 
where $\lnL_0 \equiv \ln{\left[r/(2 \rpgw) \right]}$ is the logarithmic size of GW loss cone, and $c_0$ is the constant of integration (to be deduced later numerically). For simplicity, we ignore the weak logarithmic $r$-dependence of this solution and consider the factor $\lnL_0$ in the denominator as a constant. For the case of a zero flux $\scF_p$, the above asymptotic solution is independent of $r_p$ with an explicit form $\scNi(r,r_p) = 2 N_0 r^{1-\gamma} / r_h^{3-\gamma} $ and corresponds to the so-called thermal distribution \citep{Merritt_2013}.   

Similarly at the other boundary $r_p \rightarrow 0$ (equivalent to $r_p \ll \rpgw$), we can somewhat constrain the form of solution $\scNgw(r,r_p)$ in the GW-dominated regime. By neglecting the diffusion flux ($\scF_p =0$), the steady-state FP equation~(\ref{FP_r_rp}) reduces to $\p \scF_r/ \p r = 0$ that gives $\scNgw \propto 1/\sqrt{r}$.

\emph{Brief layout of our approach}: 
We aim to find the steady-state solution $\scN(r,r_p)$ throughout the $\{r,r_p\}$-plane, which accurately describes the transition from the scattering dominated to the advection dominated regime. 
In \S~\ref{sec_ana_model} we present the simple analytical solutions in these two dynamical regimes.
An approximate global solution, widely used in the literature (see for example \citealt{Sari_Fragione_2019}) is obtained by sharply connecting these solutions at $\rpgw$, where both the two-body scatterings and GW emission are equally important, i.e., where both solutions are only marginally valid. In \S~\ref{sec_SA_model}, we present our semi-analytical model to obtain the global solution for the Fokker-Planck equation~(\ref{FP_r_rp}), utilizing the self-similarity of the solution near $\rpgw$.
We show, that our solution for the 2D number density $\scN$ improves over the previous simple theory by introducing two numerical coefficients $c_0$ and $c_1$. As elaborated in \S~\ref{sec_res}, $c_0$ corresponds to the non-vanishing density near $\rpgw$, and $c_1$ highlights the accurate location of transition region and its deviation from $\rpgw$. These coefficients allow us to estimate the rates of various galactic center processes such as EMRIs and QPEs.

\section{Approximate analytical Model}
\label{sec_ana_model}

In this section, we evaluate an approximate analytical steady-state solution of equation~(\ref{FP_r_rp}) for $\scNgw$ in GW-regime. Our approach closely follows \citet{Sari_Fragione_2019} and extrapolates the extreme dynamical regimes -- either only  diffusion or advection -- to the whole $\{r,r_p \}$-plane. This is motivated by the trajectories of stars well away from the loss cone boundary $\rpgw$ (but within the region of validity). Figure~\ref{fig_schematic} shows the schematic of the evolution of a typical stellar orbit that ends up being an EMRI. Above the loss cone boundary $r_p \gg \rpgw$, diffusion in $r_p$ due to 2B scatterings dominates and orbit evolves almost vertically \footnote{Note that the relaxation in $r$ becomes important for large $r_p$ only on longer timescales $\Tbb$.}. While well inside the boundary $r_p \ll \rpgw$, energy losses due to GWs dominate and orbit evolves almost horizontally in $\{r,r_p \}$-plane. The motion near the boundary $\rpgw$ lies in an intermediate regime where both scatterings and GW emission significantly contribute towards orbital evolution. We account for this intricate nature of dynamics in our semi-analytical model in section~\ref{sec_SA_model}, while here for the analytical model, we extrapolate the asymptotic solutions (from the two extreme regimes) till the separating boundary $r_p = \rpgw$ to cover the entire $\{r,r_p \}$-plane.   

Hence, we neglect GW emission ($\scF_r =0$) in the diffusive regime outside the loss cone ($r_p \geq \rpgw$), and similarly, neglect two-body scatterings ($\scF_p =0$) in the GW-dominated regime ($r_p < \rpgw$).  In the light of discussion on boundary conditions in section~\ref{sec_FP_formal}, the approximate steady-state solution is given as:
\beq 
\scN(r,r_p) = \begin{cases} \scNi(r,r_p) \; \mbox{; $r_p \geq \rpgw(r)$} \\[1ex] 
                    \scNgw(r,r_p) \; \mbox{; $r_p < \rpgw(r)$  }
              \end{cases}      
\eeq 
with the \emph{initial} $\scNi$ (high angular momentum boundary condition) given by equation~(\ref{scNi}). 

In this simplified picture, the phase flow is directed precisely along $r_p$ in the diffusive limit, with a $r_p$-independent flux $\scF_p$ (from equation~\ref{fluxes_Fp_Fr}) for the logarithmic solution $\scNi$: 
\beq 
\begin{split} 
\scF_p(r_{i}) &= \frac{r_h^{\gamf - 3/2}}{2 \, \Tbb(r_h)}  r_i^{5/2 - \gamf} r_p \frac{\rmd \scN_i}{\rmd r_p}  \\[1ex]
 &= \frac{1}{\lnL_0 + c_0} \; \frac{N_0}{ r_h \,  \Tbb(r_h)} \bigg( \frac{r_i}{r_h} \bigg)^{7/2 -\gamma - \gamf}
\end{split}
\label{Fp_approx}
\eeq 
where $r_i$ is an arbitrary initial semi-major axis of subject stars, that remains fixed outside the loss cone owing to the simplified assumptions of this model.  

After the star hits the loss cone boundary $r_p = \rpgw(r_i)$ and enters the GW regime, it retains this fixed value of periapsis (due to the assumption $\scF_p = 0$) while undergoing energy losses due to GW emission. In this picture, the \emph{final} periapsis $\rpf = \rpgw(r_i)$ inside the GW regime, for subject stars with an initial $r = r_i$. Hence, the net flux $\scF_r$ is directed along $r$ with the explicit form (from equation~\ref{fluxes_Fp_Fr}):  
\beq 
\scF_r(\rpf) =  \frac{r_h}{\Tgw(r_h)} \sqrt{\frac{r}{r_h}} \, \bigg( \frac{r_h}{\rpf} \bigg)^{7/2} \scNgw(r,\rpf) 
\label{Fr_approx}
\eeq
Again we know $\scNgw \propto 1/\sqrt{r}$ from boundary conditions discussed earlier, which implies $r$-independence of $\scF_r$.

Then, we can evaluate the solution $\scNgw$ by employing steady-state flow of subject stars along the limiting trajectories $r = r_i$ (in diffusive regime) and $r_p =\rpf = \rpgw(r_i)$ (in GW regime). This explicitly translates to the condition $\scF_r (\rpf) \rmd \rpf = -\scF_{p}(r_i) \rmd r_i$, which upon employing equations~(\ref{Fp_approx}) and (\ref{Fr_approx}) gives the following form of $\scNgw(r,r_p)$:  
\beq 
\begin{split} 
& \scNgw \!=\!  \frac{c_1}{\lnL_0 + c_0} \, \frac{5 }{2(3-\gamf)} \frac{2 N_0}{r_h^{2}} \bigg(\! \frac{\Tgw(r_h)}{2 \Tbb(r_h)}\!  \bigg)^{\!\! 2 \beta/5}\!\!\! \sqrt{\frac{r_h}{r}} \bigg(\! \frac{r_p}{r_h} \! \bigg)^{\beta}         \\[1em]
& = \frac{c_1}{\lnL_0 + c_0} \, \frac{5 }{2(3-\gamf)} \frac{2 N_0}{r_h^{2}} \bigg(\!  \frac{\lnL \, \mf^2 \, \Nfo}{\xi \, m \, \SMbh}  \!\bigg)^{\!\! 2 \beta/5} \!\!\!  \sqrt{\frac{r_h}{r}} \bigg(\! \frac{r_p}{R_s} \! \bigg)^{\beta} \\[1em]
&  \mbox{for } \quad  \beta(\gamma,\gamf) = \frac{5 (\gamma - 3/2)}{2(3 - \gamf)}  
\end{split} 
\label{Ngw_approx}
\eeq 
where we have deduced $r_i(\rpf)$ from $\rpf = \rpgw(r_i)$ of equation~(\ref{rp0}) and then, simplified the notation by replacing $\rpf \rightarrow r_p$. Also, we have introduced a constant prefactor $c_1$ in the above expression, which is unity for the above analytical calculation. In reality, there is a finite spreading of final periapsis around $\rpgw(r_i)$, which can lead to a non-trivial value of $c_1$. We will check the validity and consistency of this simplified analytical model (and also deduce the numerical constants $c_0$ and $c_1$), with a more general approach in the next section. 

\section{Semi-analytical model}
\label{sec_SA_model}

Here, we introduce our semi-analytical approach to find a general steady-state solution $\scN(r,r_p)$ of the FP equation~(\ref{FP_r_rp}), valid in the entire $\{r,r_p \}$-plane including both regimes of dynamics. As described in \S~\ref{sec_FP_formal}, the nature of dynamics (and hence the solution) depends on the relative location of $r_p$ with respect to the GW loss cone boundary $\rpgw(r)$. This motivates the choice of $x = r_p/\rpgw(r)$ as a preferred variable over $r_p$, with conditions -- $x \geq 1$ for diffusive and $x < 1$ for GW regime. In fact, the FP equation~(\ref{FP_r_rp}) becomes scale-free in the variables $\{r,x \}$. Further, we expect the solution $\scN$ to be self-similar for the smooth \emph{initial} distribution $\scN_i$ of subject stars. This is equivalent to considering the steady-state solution $\scN$ as separable in $r$ and $x$. Hence, by utilizing the form of asymptotic solution $\scNi$ in the diffusive regime from equation~(\ref{scNi}), we postulate the following form of general solution:
\beq 
\scN(r,r_p) = \frac{2 N_0}{r_h^{3-\gamma}} \frac{r^{1-\gamma} }{ \lnL_0 + c_0 } g(x)  
\label{N_formal} 
\eeq 
where $g$ is a positive-definite function of $x$. Substituting the above expression in the steady-state version of the FP equation~(\ref{FP_r_rp}) in transformed variables $\{r,x\}$, we get the following ordinary differential equation (ODE) in the function $g(x)$: 
\beq 
\frac{5 \, x^{\frac{5}{2} - \beta } }{2 (3 - \gamf) } \frac{\rmd}{\rmd x} \bigg( x \frac{\rmd g}{\rmd x} \bigg) + \frac{\rmd}{\rmd x} \bigg( \frac{g}{x^{\beta}}  \bigg) = 0  
\label{ode_in_g}
\eeq 
where $\beta$ is defined by equation~(\ref{Ngw_approx}). The first term dominates in diffusive regime for $x \gg 1$, while the second term dominates in the GW regime well inside loss cone for $x \ll 1$. Therefore, one can deduce the functional forms of asymptotic solutions at the extreme ends of these two regimes. In the diffusive limit $x \gg 1$, the asymptotic solution $g_i(x) \propto \log{x}$. Further, comparison between the asymptotic solution $\scN_i$ and the general solution $\scN$ from equations~(\ref{scNi}) and (\ref{N_formal}), gives $g_i(x) = \log{x} + c_0$ for the case of a constant non-zero flux $\scF_p$. But for a zero-diffusive flux condition, the asymptotic solution $g_i(x)= \lnL_0 + c_0 \, =$ constant; see the discussion below the equation~(\ref{scNi}). 

Similarly, the other asymptotic solution of equation~(\ref{ode_in_g}) in GW regime ($x \ll 1$) is of the form $\ggw(x) \propto x^{\beta}$. This validates the functional form of the asymptotic solution $\scNgw$ in equation~(\ref{Ngw_approx}) suggested by the analytical model of section~\ref{sec_ana_model}. Comparing $\scNgw$ and $\scN$ of equations~(\ref{Ngw_approx}) and (\ref{N_formal}), we have the following form of function $g$ in the GW regime: 
\beq 
g_{\rm gw}(x) = \frac{5 c_1}{2 (3-\gamf)} x^{ \beta}  \,.  
\label{g_gw_gen}
\eeq 

With the above known forms of solutions $g_i$ and $\ggw$ at the two boundaries (still with unknown coefficients $c_0$ and $c_1$), our aim is to: (1) solve for general solution $g(x)$ of the ODE of equation~(\ref{ode_in_g}) in the entire dynamical range of interest in $x$, and then (2) deduce the coefficients $\{c_0,c_1 \}$ in the asymptotic solutions. The ODE allows for an analytical solution for $\beta = 0, \, 5/2 $, as derived in appendix~\ref{app_ana_sol1}. For a BW profile of field stars with $\gamf = 7/4$, these cases correspond to $\gamma = 3/2 , \, 11/4$ for the density profile of subject stars. 

Henceforth, we specialize to the BW profile of field stars and fix $\gamf = 7/4$, for which $\beta = 2 \gamma - 3$ and the ODE of equation~(\ref{ode_in_g}) becomes: 
\beq 
{2 \, x^{{11}/{2} - 2 \gamma } } \frac{\rmd}{\rmd x} \bigg( x \frac{\rmd g}{\rmd x} \bigg) + \frac{\rmd}{\rmd x} \bigg( \frac{g}{x^{2 \gamma - 3}}  \bigg) = 0  
\label{ode_g_special}
\eeq 
As earlier, we have the asymptotic solutions: (1) for the diffusive regime with $x \gg 1$,  $g_i(x) = \log{x} + c_0$ (non-zero $\scF_p$ solution) and $g_i(x) = \lnL_0 + c_0$ (zero $\scF_p$ solution), and (2) for GW regime with $x \ll 1$, $g_{\rm gw}(x) = 2 c_1 x^{2 \gamma - 3} $ from equation~(\ref{g_gw_gen}).  

To find the final number density $\scNgw$ (and hence the corresponding flux $\scF_r$) of subject particles in GW regime (equation~\ref{Ngw_approx}) that eventually leads to observable EMRIs, we need to deduce the coefficients $\{c_1,c_0\}$ by solving the above ODE for $g(x)$. We numerically solve the ODE~(\ref{ode_g_special}) for $\gamma \in [1.505,6.5]$, which covers a wide range of physically plausible density slopes of subject stars \footnote{Note that the density slopes $\gamma \geq 3$ are possible only locally in the outer regions of a shallower density profile. }; see appendix~\ref{app_num_meth} for details.

\section{Results}
\label{sec_res}

In this section, we present the results of our semi-analytical model for the complete solution of ODE~(\ref{ode_g_special}). From the numerical solutions presented below, we find that the desired functional forms  $g_i(x)$ and $\ggw(x)$ at both boundaries are satisfied only for $\gamma \geq 3/2$ \footnote{This comment is independent of the profile of field stars; $\beta$ vanishes for $\gamma = 3/2$ for all $\gamf$ (see equation~\ref{Ngw_approx}). The desired form of boundary condition (at low $x$) $g_{\rm gw}(x)$ leads to a change in the direction of flux $\scF_p$ in GW regime at $\gamma = 3/2$. For shallower profiles of subject stars with $\gamma < 3/2$, $\scF_p$ (for $x \ll 1$) is directed towards larger $x$. In contrast for diffusive regime ($x \gg 1$), $\scF_p$ is always directed towards smaller $x$ owing to the form of $g_{i}(x)$. We believe that these oppositely directed fluxes in $r_p$ at both boundaries are responsible for the non-existence of a steady-state solution of the desired form for $\gamma < 3/2$.}. The ODE is analytically solvable for $\gamma = 3/2$ (see appendix~\ref{app_ana_sol1}) and does not permit a physical solution ($g >0$ in the entire range of $x$) for a constant non-zero $\scF_p$ condition. 
Rather, the only permissible solution for this case, corresponds to a zero $\scF_p$ condition, giving $g(x)=$ constant for all $x$. We discuss this special case and its physical implications later in section~\ref{sec_gamma_3/2}.

\begin{figure}
    \includegraphics[width=0.38\textwidth]{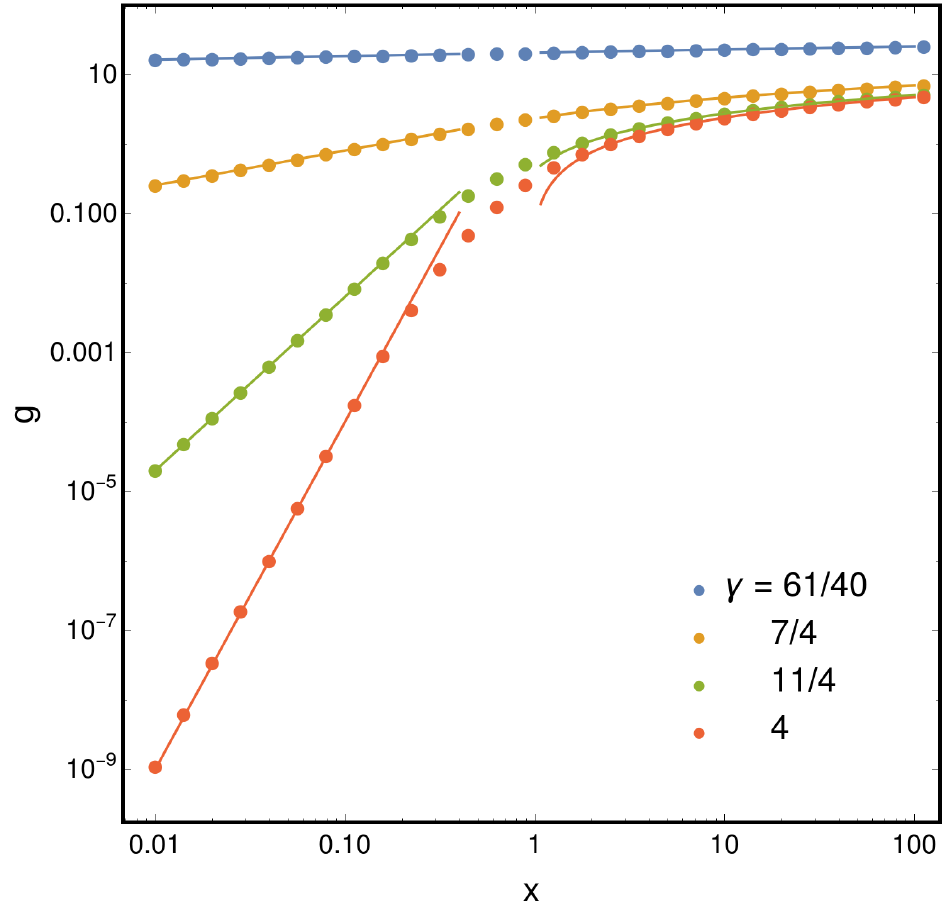}
    \caption{The numerical solution $g(x)$ (data points) for various values of the density index $\gamma$ (in color) for subject stars. The asymptotic solutions $g_i = c_0 + \log{x}$ (for $x \gg 1$) and $\ggw = 2 c_1 x^{2 \gamma -3}$ (for $x \ll 1$) are shown as solid lines in the respective regions of their validity.  }
    \label{fig_g_sol}
\end{figure}

\begin{figure*}
    \centering
    \includegraphics[width=0.8\textwidth]{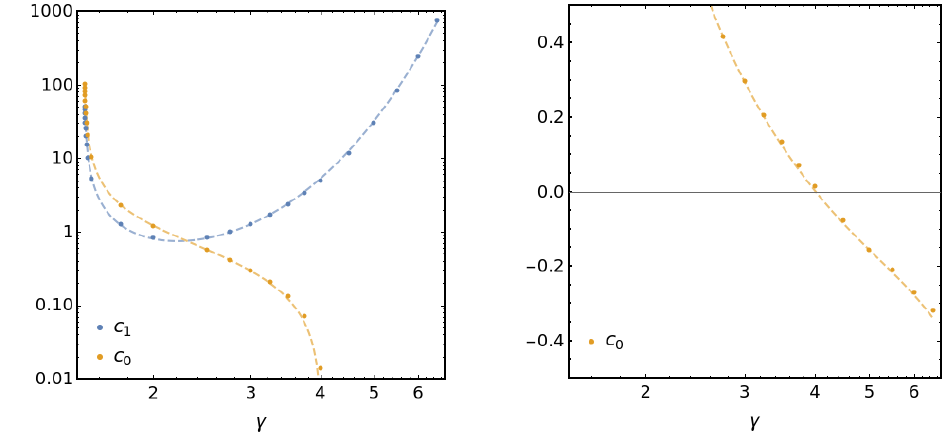}
    \caption{The numerical coefficients $c_1$ (in blue) and $c_0$ (in yellow) that determine the numerical steady-state solution are  presented as functions of density index $\gamma$ for subject stars. The points are obtained by numerically solving ODE (equation~\ref{ode_g_special}) for various $\gamma$ and dashed lines are fitting functions provided in the text. The right panel covers the negative values of $c_0$ for large $\gamma$. } 
    \label{fig_Factors}
\end{figure*}

In  figure~\ref{fig_g_sol}, we present the solution for a few representative values of $\gamma$. These solutions satisfy the desired functional forms ($g_{\rm gw}$ and $g_i$) at both boundaries. By comparing the numerical solutions with these analytical functions at both boundaries, we deduce the required coefficients $\{c_1,c_0 \}$ (see equations~\ref{c1_c0_num}). Then, we present these numerically deduced $\{c_1,c_0 \}$ as functions of $\gamma$ in figure~\ref{fig_Factors}. The numerical values for these coefficients for a few representative $\gamma$, are also provided in table~\ref{tbl_Factors}. Both these coefficients, the density prefactor $c_1$ and normalization constant $c_0$ \footnote{This terminology is motivated by the form of occurrence of these coefficients in the expressions for $\scNi$ and $\scNgw$ in equations~(\ref{scNi}) and (\ref{Ngw_approx}).}, apparently diverge to $+\infty$ as $\gamma \rightarrow 3/2^{+}$. Further, $c_0$ always decreases monotonically with increasing $\gamma$ and becomes negative for $\gamma \gtrsim 4$. For $\gamma < 11/4$, $c_1$ also decreases with increasing $\gamma$ and reverses this behavior for large $\gamma > 11/4$. Notably, $c_1 = 1$ for $\gamma = 11/4$ (see equation~\ref{c1_c0_gamma11/4}), for which the ODE~(\ref{ode_g_special}) offers an analytical solution described in appendix~\ref{app_ana_sol1}. Interestingly, the density prefactor $c_1$ remains close to unity (as suggested by the analytical model of \S~\ref{sec_ana_model}) for intermediate values of $\gamma$, with $c_1 = 1.27$ for $\gamma = 7/4$ corresponding to BW cusp of subject stars. Hence, the results are expected to deviate significantly ($c_1 \gtrsim 2$) from the analytical model only for either very small $\gamma \lesssim 1.7$ or large $\gamma \gtrsim 3.4$. We explain physically these properties of $\{c_1,c_0 \}$ in the later part of this section~\ref{sec_c1} and \ref{sec_c0}.

\begin{table}[t!]
\centering
\footnotesize
\begin{tabular}{|c| c|c| c|}
  \hline
$\gamma$  & $c_1$ & $c_0$ & $c_p$  \\
  \hline 
121/80 & 20.2   &   40.4   &  3.37    \\[1em] 
  
  61/40  & 10.2 &  20.4    & 2.58    \\[1em] 
  
  31/20  & 5.22 &  10.4    & 1.99        \\[1em]  

  7/4  & 1.27 &  2.30   & 1.13   \\[1em]  


 11/4  & 1 & 0.415  & 0.637 \footnote{For this case, $c_p$ is deduced from the fitting function of equation~(\ref{cp_fitting_func}).}    \\[1em] 
  
  4   & 5.0   & 0.0139  & 0.525      \\[1em]  

  13/2 & 750 & -0.320 &  0.414    \\[1ex] 

  \hline 

\end{tabular}
\caption{The coefficients $\{c_1, c_0\}$ and associated periapsis prefactor $c_p$ for the numerical solution are given for a set of representative values of $\gamma$. These values of $\gamma$ (downwards) are chosen so that they correspond to a thermally relaxed state in energy, with $\gamma = 3/2 + m/(4 \mf)$ \citep{Bahcall_Wolf_1977} for the mass ratio $m/\mf = \{ 1/20,1/10,1/5,1,5,10,20 \}$.  }
\label{tbl_Factors}
\end{table}

Here we provide the fitting functions for $c_1$ and $c_0$, with an accuracy of $\simeq 10\%$ for $\gamma \in [1.505, 6.5]$: 
\begin{subequations} 
\begin{align}  
& c_1 = \bigg(  \frac{  0.57 }{(\gamma - 3/2)^{2/5}} +0.11   \bigg)^{11/2 - 2 \gamma}  \\[1ex]
& c_0 =    \frac{0.5}{( \gamma - 3/2)} + 1.2 - 0.7 \gamma + 0.12 \gamma^2 - 0.0073 \gamma^3 
\end{align}
\label{fitting_func_c1_c0}
\end{subequations} 
From figure~\ref{fig_Factors}, it is evident that these functions (dashed curves) follow quite well the variation of numerically computed $\{c_1,c_0 \}$ (data points) throughout the explored range of $\gamma$. 

\subsection{Effective final periapsis $\rpf$ }
\label{sec_cp}

\begin{figure*}[t!]
    \centering
    \includegraphics[width=0.98\textwidth]{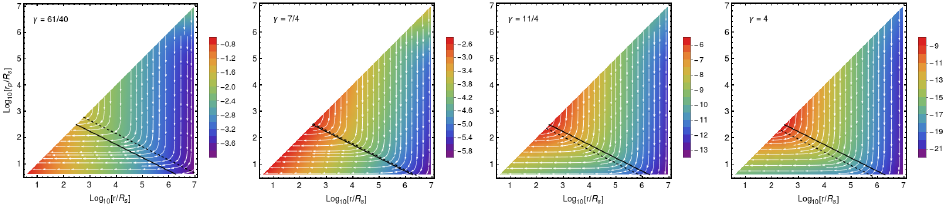}
    \caption{The 2D number density $\scN$ (in color) of subject stars in $\{r,r_p\}$-plane for $\gamma = \{61/40,7/4,11/4,4 \}$ (increasing towards right). The color represents the quantity $\log_{10}[\scN/(2 N_0 R_s^{1-\gamma}/r_h^{3-\gamma})]$. The streamlines (as white arrows) indicate the direction of net flux. The thick black line represents the GW \emph{loss cone} boundary $r_p = \rpgw(r)$ (equation~\ref{rp0}). The dashed black line indicates the effective perispasis $\rpf = c_p \rpgw$, where the direction of net flux (or streamlines) effectively change their behavior. For smaller $\gamma \lesssim 7/4$, the transition occurs at higher $\rpf > \rpgw$; while $ \rpf < \rpgw$ for high values of $\gamma$. Here we consider BHs ($\mf = 10\Msun$) as subject stars being scattered by background stars ($\mf = 1 \Msun$), with mass $\SMbh = 10^6 \Msun$ for central MBH. Also, the reference radius $r_h =  9.1 \times 10^6 R_s$ ($r_h$ being the radius of influence of MBH), $\Nfo = 10^6$, $\lnL = 10$ and $\lnL_0 = 10$.   } 
    \label{fig_streams}
\end{figure*}

It is instructive to visualize the direction of the net flow of subject stars in $\{r,r_p \}$-plane.  
 We evaluate 2D density $\scN(r,r_p)$ (equation~\ref{N_formal}) and fluxes $\{ \scF_r, \scF_p \}$ (equation~\ref{fluxes_Fp_Fr}) by employing the numerically computed solution $g(x)$ above, for some typical values of $\gamma = \{61/40,7/4,11/4,4 \}$. Figure~\ref{fig_streams} presents the 2D density $\scN$ maps (in color), with streamlines (in white arrows) showing the direction of local net flux. The local slope of a streamline is ${\rmd {r_p} }/{ \rmd r} = \scF_p/\scF_r =  ({r_p}/{r}) {x^{5/2}} {\rmd \log{g}}/{\rmd \log{x} } $. The vertically aligned net flux (along $r_p$) indicates the dominant contribution from diffusion due to 2B scatterings, while effectively horizontal net flux (along $r$) indicates the dominance of energy losses due to GW emission. Naively, one would expect the transition of net fluxes - from vertical to horizontal - to occur near the GW loss cone boundary $\rpgw(r)$ (equation~\ref{rp0}; shown as a black solid line in the figure). The figure shows that this transition actually occurs at an effective periapsis $\rpf = c_p \rpgw$ (shown as a dashed black line),  which increases for smaller values of $\gamma$. Hence, the coefficient $c_p$ is a monotonically decreasing function of $\gamma$. 

Note that the transition in the nature of individual subject star orbits (which are independent of the density slope $\gamma$, but only depend on density slope $\gamf$ of field stars) always occurs near $r_p = \rpgw$. Hence we expect that the number density profiles for small $\gamma$, become nearly uniform in $r_p$ for $r_p \lesssim \rpf$, so that net diffusive flux $\scF_p$ becomes extremely small even in a portion of the diffusive domain above $\rpgw$. This enhancement in the number density can be due to incoming subject stars diffusing from larger initial $r$, and is favored for shallow cusps with small $\gamma$ hosting a relatively higher number of stars at larger $r$.

 We can relate this periapsis prefactor $c_p$ to the density prefactor $c_1$, because modulation of final density $\scNgw$ by introducing prefactor $c_1$ (as done in \S~\ref{sec_ana_model}) is equivalent to correcting the final periapsis $\rpf$ in GW regime by employing $c_p$. This relation is determined from the condition that the advective flow at the \emph{final} $\rpf$ in GW regime $\scF_r(\rpf) \rmd \rpf \propto c_1 \rpf^{\beta} \rmd \rpf/\rpf^{7/2} = c_1 c_p^{\beta-5/2} $ (from equations~\ref{Fr_approx} and \ref{Ngw_approx}), is equal to the diffusive flow $\scF_p(r_i) \rmd r_i$ at an \emph{initial} $r = r_i$ (for $ x \gg 1$). This implies that the net flow $\scF_r(\rpf) \rmd \rpf$ remains the same for both analytical and numerical solutions. This gives $c_p = c_1^{-1/(\beta - 5/2)}$, and for the special case of interest with a BW profile of the field stars, we have $c_p = c_1^{-1/(2 \gamma - 11/2)}$. Using this relation, we calculate $c_p$ for some values of $\gamma$, as quoted in table~\ref{tbl_Factors}.  Employing the fitting function for $c_1$ (equation~\ref{fitting_func_c1_c0}) from numerical solutions, we have the following explicit form of $c_p$ as a function of $\gamma$: 
 \beq
c_p =   \frac{  0.57 }{(\gamma - 3/2)^{2/5}}  + 0.11 \,
\label{cp_fitting_func}
 \eeq 
As expected from the above physical arguments, $c_p$ is a monotonically decreasing function of $\gamma$. This dependence can also be appreciated geometrically from the fact that $\rpf$ or $c_p$ depends on the shape of flux tubes (figure~\ref{fig_streams}). The slope of net flux streamlines $\rmd r_p/ \rmd r \propto \rmd \log{g}/\rmd \log{x} = 2 (\gamma - 3/2)$ in the GW regime (see discussion below equation~\ref{ode_g_special}) and the higher slope implies the lower final periapsis $\rpf$. This explains the inverse dependence of $c_p$ on $(\gamma- 3/2)$.   

\subsection{Density prefactor $c_1$} 
\label{sec_c1}

As evident from the above discussion, the density prefactor $c_1$ essentially locates the transition boundary between the extreme dynamical regimes and its deviation from the loss cone boundary $\rpgw$. Given the relation $c_1 = c_p^{11/2 - 2 \gamma}$, it is straightforward to explain various properties of $c_1$, as illustrated in figure~\ref{fig_Factors}:
\begin{itemize}
\item[(a).] $c_1 = 1$ for $\gamma = 11/4$.

\item[(b).] For small $\gamma < 11/4$, $c_1$ behaves similar to $c_p$ and increases with decreasing $\gamma$.

\item[(c).] For large $\gamma > 11/4$, $c_1$ has an inverse relation to $c_p$ and hence, becomes an increasing function of $\gamma$. 
\end{itemize}

Now, we explain physically the above mathematical properties.  
In GW regime ($x \ll 1$), the advective flow  $\scF_r(r_p) r_p \propto \scNgw  r_p/\Tgw \propto  r_p^{2( \gamma - 11/4)}$, is channeled by the diffusive flow $\scF_p(r_i) r_i \propto r_i^{11/4-\gamma}$ (from equation~\ref{Fp_approx}) at the initial $r_i$ in the diffusive limit ($x \gg 1$). For $\gamma < 11/4$, larger $r_i$ (smaller $\rpf$) contribute higher diffusive (advective) flow. As seen in \S~\ref{sec_cp}, the effective transition boundary $\rpf$ is a monotonically decreasing function of $\gamma$. As $\gamma$ decreases for increasingly shallower profiles, the advective flow at a given $r_p$ is supplied by subject stars diffusing from higher initial $r_i$. This demands an increase in advective flow over usual estimates of the analytical model, which is corrected by higher values of $c_1$. This explains the decreasing behavior of $c_1$ with $\gamma$ for $\gamma < 11/4$. Similarly for increasingly steep profiles with $\gamma > 11/4$, advective flow at a given $r_p$ is supplied by smaller $r_i$, which have higher diffusive flow for $\gamma > 11/4$. This explains the increasing behavior of $c_1$ with $\gamma$ for $\gamma > 11/4$. Notably, for $\gamma = 11/4$, all initial $r_i$ contribute equal flow $\scF_p(r_i) r_i$ in the diffusive regime, and no correction is needed over the analytical model leading to $c_1= 1$.

\subsection{Normalization constant $c_0$} 
\label{sec_c0}

 As argued in \S~\ref{sec_cp}, an enhancement in number density $\scN$ near loss cone boundary $r_p = \rpgw$ is expected for shallow cusps with small $\gamma$. This happens due to a fraction of incoming subject stars from larger initial $r_i$, that diffuse to larger $r_p$ along the loss cone boundary. Increase in $c_0$ for smaller $\gamma$ (figure~\ref{fig_Factors}), captures this density enhancement as $\scN_i \propto (\log{x}+c_0)/(\lnL_0 + c_0)$ increases with $c_0$.\footnote{The denominator ensures that the number density for nearly circular orbits ($r_p \simeq r$) does not change, due to longer relaxation timescales. Note that $\lnL_0 = \log{(r/\rpgw)}$ for some typical $r$ of interest, but we treat this quantity as independent of $r$ for simplicity.  }

\subsection{Shallow profiles -- $\gamma \rightarrow 3/2^{+}$ case  } 
\label{sec_gamma_3/2}

It is important to address the case of small $\gamma$ separately, as both the numerical coefficients $\{c_1,c_0\}$ deduced from numerical solutions tend to diverge to $+\infty$ as $\gamma \rightarrow 3/2^{+}$. Now, we check the physical interpretation of this divergence and its implications for number density and fluxes in the entire $\{ r,r_p\}$-plane. 

In the diffusive regime, number density $\scN_i \propto (\log{x}+c_0)/(\lnL_0 + c_0)$ from equation~(\ref{scNi}), giving a uniform profile in $r_p$ as $c_0 \rightarrow+\infty$. Hence, the diffusive flux $\scF_p \propto r_p \p \scN_i/\p r_p \propto 1/(\lnL_0 + c_0) $ $\rightarrow 0$ for diverging $c_0$. In the GW regime, we have the number density $\scNgw \propto g_{\rm gw}(x)/(\lnL_0 + c_0) \propto 2 c_1/c_0$ from equations~(\ref{Ngw_approx}), (\ref{N_formal}) and (\ref{g_gw_gen}). Evidently, the dependence of $\scNgw$ on $r_p$ or $x$ vanishes for $\gamma = 3/2$ (similar to the asymptotic solution $\scN_i$ in the diffusive regime). We can deduce the limiting factor $(2 c_1/c_0)$ by using the fitting functions for $\{c_1,c_0 \}$ from equation~(\ref{fitting_func_c1_c0}). Indeed, $2 c_1 / c_0 \rightarrow 1$ as $\gamma \rightarrow 3/2$, inspite of individual divergence of $c_0$ and $c_1$. Hence, the density profile, written explicitly as (from equation~\ref{N_formal}):
\beq 
\scN = \frac{2 N_0}{ r_h^{3/2}} \frac{1}{\sqrt{r}} 
\label{N_gamma_3/2}
\eeq 
is independent of $r_p$ in the entire $\{r,r_p \}$-plane and exactly matches the initial profile $\scNi$ for a zero $\scF_p$ solution (see the discussion below equation~\ref{scNi}). This leads to a zero diffusive flux $\scF_p$, and a uniform advective flux $\scF_r$ (from equation~\ref{fluxes_Fp_Fr}):
\beq 
\scF_r =    \frac{2 N_0}{r_h \Tgw(r_h)} \bigg( \frac{r_h}{r_p} \bigg)^{7/2} \,. 
\label{Fr_gamma_3/2}
\eeq 
The limiting solution derived above exactly matches the analytical solution for this case with $\gamma = 3/2$, presented in appendix~\ref{app_ana_sol1}. So, the effect of divergence of $c_1$ and $c_0$ for shallow profiles is to suppress the logarithmic damping factor $1/\lnL_0 $ (arising from the 2D nature of angular momentum relaxation) in number density and advective flux in the GW regime, because $2 c_1/(\lnL_0 + c_0) \rightarrow 1$ as $\gamma \rightarrow 3/2^{+}$. In comparison to the analytical model, our calculation would predict an enhanced rate of EMRI formation for shallow profiles, amplified upto the order of $\lnL_0 \equiv \log{(r_i/\rpgw)}$. We quantify this enhancement in EMRI rates in \S~\ref{sec_discus}.

\subsection{Steep profiles - high $\gamma$ case} 
\label{sec_steep_gamma}

For steep profiles with $\gamma > 11/4$, $c_1$ increases, while $c_0$ decreases with increasing $\gamma$. This implies amplified number density $\scNgw$ and advective flux $\scF_r$ of subject stars in GW regime, that would eventually become EMRIs. But, as noted earlier, a significant amplification ($c_1 \gtrsim 2$) occurs for only very steep profiles with $\gamma \gtrsim 3.4$. But, it is possible to have such steep profiles only locally because $\gamma < 3$ is required for a global profile. So, the actual amplification in EMRI rates contributed by high $\gamma$ regions might be only limited.

\section{Discussion and Conclusions}
\label{sec_discus}

In this work, we consider a two-population model of an NSC around a central MBH, hosting subject stars immersed in a background cusp of field stars, which act as dominant scatterers. In our Fokker-Planck framework, we include : {(1)~angular} momentum relaxation owing to scatterings among subject stars over time $ \TL$, and {(2)~energy} losses from GW emission over time $\Tgw$ due to interaction with the central MBH. Due to the hierarchy of these timescales, stellar dynamics can be essentially separated into two regimes where only one of these effects dominates in the extreme limits, and one can deduce analytical asymptotic solutions. 

We denote the low-angular momentum region with periapse $r_p \leq \rpgw(r)$ as the GW loss cone. Both angular momentum diffusion and energy advection are equally important near $\rpgw$, where the transition between the two regimes occurs. Employing the asymptotic solutions and suitable boundary conditions, we numerically solve for the steady-state solutions of the Fokker-Planck equation for 2D number density $\scN$ of subject stars in the entire $\{r,r_p\}$-plane. Our semi-analytical approach aims at accurately treating the region near loss cone boundary $\rpgw$, so that the transition from diffusion-dominated phase to GW-dominated phase is captured properly. The resulting distribution in the GW regime, $\scNgw$, represents the progenitor population that gradually evolves into observable EMRIs and hence, determines their formation rate and orbital characteristics.   

We obtain the semi-analytical steady-state solution by first transforming the problem into a physically motivated unit $x = r_p/\rpgw$ that makes the remnant FP equation in $(x,r)$ scale-free. This allows for self-similar solutions $\scN \propto r^{1-\gamma} g(x)$. This reduces the initial 2D partial differential equation~(\ref{FP_r_rp}) for $\scN(r,r_p)$ into an ODE~(\ref{ode_in_g}) for $g(x)$, which we solve numerically for a wide range of physically relevant power-law density indices $\gamma$ for subject stars, while considering a BW cusp with $\gamf = 7/4$ of the background field stars. For a few special density indices $\{\gamma,\gamf \}$, this ODE allows for analytical solutions presented in appendix~\ref{app_ana_sol1}.  

The numerical solution we find for $g(x)$ allows us to accurately connect the asymptotic regimes of 2B scattering and GW emission. In these terms, our solution provides the necessary coefficients $\{c_1,c_0\}$, which depend only on $\gamma$. This provides the number density and fluxes in the GW regime as
\begin{equation}
\begin{split}
& \scNgw \!=\!   \frac{c_1}{\lnL_0 + c_0}  \frac{5 }{2(3-\gamf)} \frac{2 N_0}{r_h^{2}} \bigg(\!  \frac{\lnL \, \mf^2 \, \Nfo}{\xi \, m \, \SMbh} \! \bigg)^{\!\!\! \frac{2 \beta}{5} } \!\!\!  \sqrt{\frac{r_h}{r}} \bigg(\! \frac{r_p}{R_s} \!  \bigg)^{\beta} \\[1ex]
&  \scF_r =  \frac{85 \, c}{96 \sqrt{2}}  \frac{m}{\SMbh}   
  \bigg( \frac{R_s}{r_p} \bigg)^{\!7/2} \, \sqrt{\frac{r}{R_s}} \;  \scNgw(r,r_p) 
\end{split}
\label{Ngw_Fr_fin}
\end{equation}
with $\beta(\gamma,\gamf) = {(5/2) (\gamma - 3/2)}/{(3 - \gamf)}$. 
We compute the numerical solution (and hence $\{c_1,c_0\}$) only for a BW cusp of field stars, but this could be easily repeated for any field population. The resulting coefficients $\{c_1,c_0\}$ are presented in figure~\ref{fig_Factors} and table~\ref{tbl_Factors}. We also provide fitting functions (equation~\ref{fitting_func_c1_c0}) for these coefficients for a broad range of $\gamma$. The coefficients $c_0$ and $c_1$ physically correspond to the non-vanishing density near the loss-cone boundary and its accurate position respectively.

We find that our semi-analytical model predicts an amplified flux $\scF_r$ over the analytical model with $c_1 \gtrsim 2$, only for either very shallow ($\gamma \lesssim 1.7$) or steep profiles ($\gamma \gtrsim 3.4$). 
While most models do not exhibit such steep profiles \citep{Hopman_2006,Keshet_2009,Fragione_Sari_2018,Sari_Fragione_2019}, they do occur for BHs in the outskirts of the sphere of influence
\citep{Linial_Sari_22,Rom_24}.
On the other hand, the enhancement of flux for shallow profiles can be physically relevant for MS EMRIs \citep{Linial_Sari_2023,Rom_24}. 

\begin{figure}
    \centering
    \includegraphics[width=0.4\textwidth]{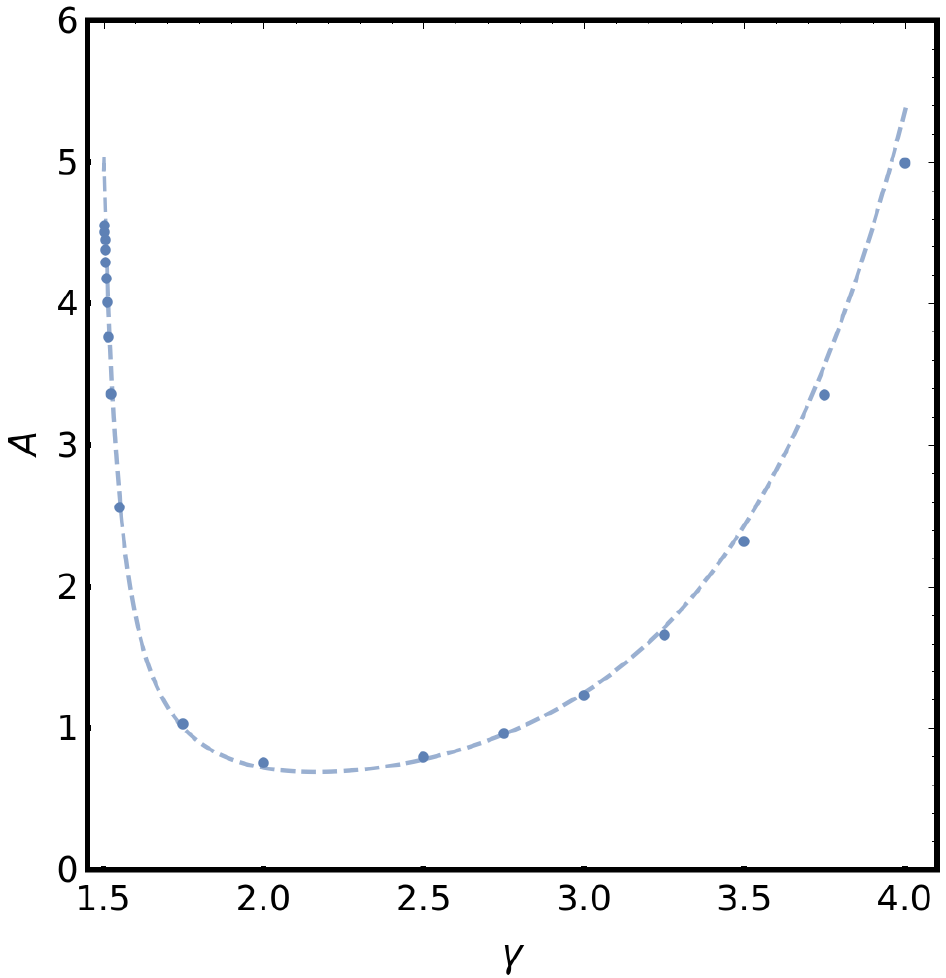}
    \caption{Amplification factor $A$, the ratio of EMRI rates predicted by semi-analytical model to naive analytical model, for a typical value of $\lnL_0 = 10$. Here $A$ is plotted as a function of $\gamma$, the density profile index of subject stars, while the field stars form a BW cusp. Data points correspond to numerical solution and dashed line is evaluated from the fitting functions for $c_1$ and $c_0$ (equation~\ref{fitting_func_c1_c0}).    } 
    \label{fig_amp_fac}
\end{figure}

\emph{EMRI rates \& Amplification factor}: 
For most of the physically relevant stellar distributions, the diffusive flux increases with the semi-major axis $r_i$, which is valid for $\gamma + \gamf < 9/2$ (from equation~\ref{Ngw_Fr_fin}). Therefore, the EMRIs formation rate corresponds to advective flow near the capture radius $\rcap$, and can be estimated as $\Remri = \scF_r(r_p = \rcap) \rcap$ \citep{Hopman_Alexander_2005}, with the following explicit expression (using equation~\ref{Ngw_Fr_fin}):
\beq 
\begin{split}
\Remri &= A\, \frac{85 c }{96\sqrt{2}\lnL_0 } \, \frac{5 }{2(3-\gamf)} \,\frac{m}{\SMbh} \,\frac{2 N_0}{r_h}  \\[1ex]
& \qquad \qquad \bigg(  \frac{\lnL \, \mf^2 \, \Nfo}{\xi \, m \, \SMbh}  \bigg)^{\!\! 2 \beta/5} \sqrt{\frac{R_s}{r_h}} \bigg( \frac{\rcap}{R_s}  \bigg)^{\beta - 5/2},  
\end{split}
\label{rates}
\eeq 
where
\beq 
A = \frac{c_1 \lnL_0}{\lnL_0 + c_0} ,
\label{amp_fac}
\eeq 
is the amplification factor, compared to the rate obtained from the analytical scheme (\S~\ref{sec_ana_model}) that simply connects the asymptotic solutions \footnote{Analytical solution corresponds to $c_0 = 0$, $c_1 = 1$, and hence $A = 1$.}.

As an example, we evaluate here the BH EMRI rate, $\Remri_{\rm BH}$, and the MS EMRI rate, $\Remri_{\rm MS}$, considering BHs following BW density profile as the dominant scatterers, with $\mf = 10 \Msun$, $\Nfo = 10^{-3} f_{-3} \SMbh/\Msun$ and $\gamf = 7/4$. This yields $\Remri_{\rm BH} = 6.2 \, {\rm Gyr}^{-1} f_{-3}^{6/5} \SMBhMW^{-1/4}$, where $\SMBhMW = \SMbh/(4 \times 10^6 \Msun)$ is normalized with respect to the Galactic center MBH mass \citep{Ghez_2008,Gillessen_2009}. Here we choose $\rcap = 4 R_s$, $\lnL = \lnL_0 = 10$, and $r_h = 2~{\rm pc}\, \sqrt{\SMBhMW}$ (as obtained from $\SMbh-\sigma$ relation, \citealt{Kormendy_Ho_2013}). This BH EMRI rate is comparable to the rates obtained by \citet{Hopman_Alexander_2005} for a similar choice of parameters. For steeper density profiles \citep{Amaro_Seoane_Preto_2011,Rom_24}, the EMRIs formation rate is higher by 1-2 orders of magnitudes. However, ejections resulting from strong encounters, which we do not consider here, may lower the EMRI rates \citep{Kaur_Perets_24} \footnote{This effect has been already seen in the context of TDE rates \citep{Teboul_2024}.}.

Figure~\ref{fig_amp_fac} shows the variation in the amplification factor $A$ as a function of subject stars density index $\gamma$ immersed in a BW cusp of field stars for a representative value of $\lnL_0 = 10$. For steeper profiles, the results vary appreciably ($A \gtrsim 2$) from the analytical model only for $\gamma \gtrsim 3.4$, that can occur locally for BHs near the outskirts of their distribution \citep{Rom_24}. For moderately steep profiles $7/4 \lesssim \gamma \lesssim 3$, the correction $A$ is close to unity. For shallow profiles with $\gamma \rightarrow 3/2^{+} $, both $c_1$ and $c_0$ diverge to $+ \infty$; while their ratio $c_1/c_0 \rightarrow 1/2$ (check \S~\ref{sec_gamma_3/2} for details), implying $A \rightarrow \lnL_0 /2$. Hence, for $\lnL_0 =10$, the EMRI rate increases by a factor of $\sim$5 for shallow profiles relevant for MS EMRIs as shown below. 

The MS EMRI rates $\Remri_{\rm MS}$ for a shallow profile of stars with $\gamma = 3/2$ (and $m = 1 \Msun$, $N_0 = \SMbh/\Msun$) and a BW profile of BHs as the dominant scatterers as earlier, yields $\Remri_{\rm MS} = 3.2 \times 10^{-7}~{\rm yr}^{-1} \SMBhMW^{17/12}$ \footnote{Here $\rcap = \rtid = 9.3~R_s \SMBhMW^{-2/3}$ for a Sun-like star.}. This rate is higher, by an order of magnitude, compared to the previous estimates for a BW cusp of stars \citep{Linial_Sari_2023}. Most of this difference arises from our refined treatment of the transition between the gravitational wave and two-body scattering regimes (i.e., the amplification factor $A$).
Notably, the presence of binaries may induce a comparable or higher contribution to the EMRI rates, via Hills mechanism \citep{Hills1988} depending on the binary fraction in galactic centers \citep{Sari_Fragione_2019,Linial_Sari_2023}. However, the formation rate of the MS EMRIs may be significantly reduced by physical collisions among stars \citep{Sari_Fragione_2019,Rose_2023,Balberg_2023}, which are not considered here.

Based on eROSITA surveys, \citet{Arcodia_2024} have recently inferred a lower limit on QPE rates $\Gamma_{\rm QPE} \geq 4 \times 10^{-6}(\tau/10 {\rm yr})^{-1}~{\rm yr}^{-1}$, with $\tau$ as the life-time of single QPE episode. 
\cite{Linial_Metzger_2023} have recently suggested that QPEs originate from the interaction of a pre-existing MS or BH EMRI with a newly formed TDE debris disk. We now test this QPE $=$EMRI$+$TDE hypothesis, from the point of view of rates. The observed QPE rate is much larger than the theoretical estimates of EMRI rates. 
Assuming that an EMRI can survive its QPE phase, this rate discrepancy requires that different TDEs trigger the same EMRI to repeatedly become a QPE (these repeated active periods as QPE are separated by roughly $10^5$ years). We therefore estimate the QPE rate as $\Gamma_{\rm QPE} = \Gamma_{\rm EMRI} \times \Gamma_{\rm TDE} \times \Tgw$. We take a TDE rate $\Gamma_{\rm TDE} \approx 10^{-5}\,{\rm yr}^{-1}$ \citep{Gezari_2009}, and life-time of an EMRI progenitor as the GW inspiral time $\Tgw \approx 2 \times 10^7~{\rm yr} \, (\Msun/m)$ for an orbital period of a day \footnote{For the estimates in this paragraph, we employ $\SMbh = 4 \times 10^6~\Msun$ corresponding to the Galactic center MBH.}. For a MS star$+$disk interaction scenario \citep{Tagawa_2023,Linial_Metzger_2023}, we use $\Remri_{\rm MS} = 3.2 \times 10^{-7}~{\rm yr}^{-1}$ and get $\Gamma_{\rm QPE} \approx 6 \times 10^{-5}~{\rm yr}^{-1}$, consistent with the observed lower limit. For a BH$+$disk scenario \citep{Franchini_2023}, we have $\Gamma_{\rm QPE} \approx 10^{-7}~f_{-3}^{6/5}~{\rm yr}^{-1}$, using the BH EMRI rates evaluated above for a BW profile. The latter scenario could become consistent with observations for a high number fraction of BHs within $r_h$, $\Nfo {(\Msun/\SMbh)}\gtrsim 0.02$ , or a longer QPE lifetime $\tau$. 

Finally, we note that since we neglect plunge/TDE loss cone, our approach does not accurately discriminate EMRIs from plunges near the critical semi-major axis $r_c$ \citep{Hopman_Alexander_2005}. Further, we consider power-law density profiles for both subject and field stars with infinite spatial range. However, stars and BHs may exchange the role of dominant scatterers in different spatial locations leading rather to broken power-law profiles \citep{Rom_24}. Hence, our results may be inaccurate near the transition.

Our approach brings out the underlying physical simplicity, in terms of self-similarity, of the scattering-driven formation of nuclear transients and provides useful improvements for their occurrence rates. Upcoming GW observatories LISA and TianQin will appraise different EMRI formation channels, while scrutinizing the predictions for rates and characteristics. Similarly, current and upcoming X-ray survey telescopes, like eROSITA will give better constrains on nature and rates of QPEs with more number of secure candidates. 
These observational advances would demand a careful assessment of theoretical models proposed for these transients.

\begin{acknowledgments}
This research was partially supported by an ISF grant, an NSF/BSF grant, and an MOS grant. B.R. acknowledges support from the Milner Foundation. K.K. gratefully acknowledges the hospitality and support of Max Planck Institute for Extraterrestrial Physics, where a part of this work was done.    
\end{acknowledgments}


\bibliography{main}{}

\begin{thebibliography}{}
\expandafter\ifx\csname natexlab\endcsname\relax\def\natexlab#1{#1}\fi
\providecommand{\url}[1]{\href{#1}{#1}}
\providecommand{\dodoi}[1]{doi:~\href{http://doi.org/#1}{\nolinkurl{#1}}}
\providecommand{\doeprint}[1]{\href{http://ascl.net/#1}{\nolinkurl{http://ascl.net/#1}}}
\providecommand{\doarXiv}[1]{\href{https://arxiv.org/abs/#1}{\nolinkurl{https://arxiv.org/abs/#1}}}

\bibitem[{{Aharon} \& {Perets}(2016)}]{Aharon_Perets_2016}
{Aharon}, D., \& {Perets}, H.~B. 2016, \apjl, 830, L1,
  \dodoi{10.3847/2041-8205/830/1/L1}

\bibitem[{{Alexander}(2017{\natexlab{a}})}]{Tal_2017}
{Alexander}, T. 2017{\natexlab{a}}, \araa, 55, 17,
  \dodoi{10.1146/annurev-astro-091916-055306}

\bibitem[{{Alexander}(2017{\natexlab{b}})}]{Alexander_2017RR}
{Alexander}, T. 2017{\natexlab{b}}, in Journal of Physics Conference Series,
  Vol. 840, Journal of Physics Conference Series, 012019,
  \dodoi{10.1088/1742-6596/840/1/012019}

\bibitem[{{Alexander} \& {Hopman}(2009)}]{Alexander_Hopman2009}
{Alexander}, T., \& {Hopman}, C. 2009, \apj, 697, 1861,
  \dodoi{10.1088/0004-637X/697/2/1861}

\bibitem[{{Amaro-Seoane}(2018)}]{Amaro-Seoane_2018}
{Amaro-Seoane}, P. 2018, Living Reviews in Relativity, 21, 4,
  \dodoi{10.1007/s41114-018-0013-8}

\bibitem[{{Amaro-Seoane} {et~al.}(2007){Amaro-Seoane}, {Gair}, {Freitag},
  {Miller}, {Mandel}, {Cutler}, \& {Babak}}]{Amaro-Seoane_2007}
{Amaro-Seoane}, P., {Gair}, J.~R., {Freitag}, M., {et~al.} 2007, Classical and
  Quantum Gravity, 24, R113, \dodoi{10.1088/0264-9381/24/17/R01}

\bibitem[{{Amaro-Seoane} \& {Preto}(2011)}]{Amaro_Seoane_Preto_2011}
{Amaro-Seoane}, P., \& {Preto}, M. 2011, Classical and Quantum Gravity, 28,
  094017, \dodoi{10.1088/0264-9381/28/9/094017}

\bibitem[{{Amaro-Seoane} {et~al.}(2012){Amaro-Seoane}, {Aoudia}, {Babak},
  {Bin{\'e}truy}, {Berti}, {Boh{\'e}}, {Caprini}, {Colpi}, {Cornish},
  {Danzmann}, {Dufaux}, {Gair}, {Jennrich}, {Jetzer}, {Klein}, {Lang}, {Lobo},
  {Littenberg}, {McWilliams}, {Nelemans}, {Petiteau}, {Porter}, {Schutz},
  {Sesana}, {Stebbins}, {Sumner}, {Vallisneri}, {Vitale}, {Volonteri}, \&
  {Ward}}]{Amaro-Seoane_2012lisa}
{Amaro-Seoane}, P., {Aoudia}, S., {Babak}, S., {et~al.} 2012, Classical and
  Quantum Gravity, 29, 124016, \dodoi{10.1088/0264-9381/29/12/124016}

\bibitem[{{Amaro-Seoane} {et~al.}(2017){Amaro-Seoane}, {Audley}, {Babak},
  {Baker}, {Barausse}, {Bender}, {Berti}, {Binetruy}, {Born}, {Bortoluzzi},
  {Camp}, {Caprini}, {Cardoso}, {Colpi}, {Conklin}, {Cornish}, {Cutler},
  {Danzmann}, {Dolesi}, {Ferraioli}, {Ferroni}, {Fitzsimons}, {Gair}, {Gesa
  Bote}, {Giardini}, {Gibert}, {Grimani}, {Halloin}, {Heinzel}, {Hertog},
  {Hewitson}, {Holley-Bockelmann}, {Hollington}, {Hueller}, {Inchauspe},
  {Jetzer}, {Karnesis}, {Killow}, {Klein}, {Klipstein}, {Korsakova}, {Larson},
  {Livas}, {Lloro}, {Man}, {Mance}, {Martino}, {Mateos}, {McKenzie},
  {McWilliams}, {Miller}, {Mueller}, {Nardini}, {Nelemans}, {Nofrarias},
  {Petiteau}, {Pivato}, {Plagnol}, {Porter}, {Reiche}, {Robertson},
  {Robertson}, {Rossi}, {Russano}, {Schutz}, {Sesana}, {Shoemaker}, {Slutsky},
  {Sopuerta}, {Sumner}, {Tamanini}, {Thorpe}, {Troebs}, {Vallisneri},
  {Vecchio}, {Vetrugno}, {Vitale}, {Volonteri}, {Wanner}, {Ward}, {Wass},
  {Weber}, {Ziemer}, \& {Zweifel}}]{Amaro-Seoane_2017}
{Amaro-Seoane}, P., {Audley}, H., {Babak}, S., {et~al.} 2017, arXiv e-prints,
  arXiv:1702.00786, \dodoi{10.48550/arXiv.1702.00786}

\bibitem[{{Arcodia} {et~al.}(2021){Arcodia}, {Merloni}, {Nandra}, {Buchner},
  {Salvato}, {Pasham}, {Remillard}, {Comparat}, {Lamer}, {Ponti}, {Malyali},
  {Wolf}, {Arzoumanian}, {Bogensberger}, {Buckley}, {Gendreau}, {Gromadzki},
  {Kara}, {Krumpe}, {Markwardt}, {Ramos-Ceja}, {Rau}, {Schramm}, \&
  {Schwope}}]{Arcodia_2021}
{Arcodia}, R., {Merloni}, A., {Nandra}, K., {et~al.} 2021, \nat, 592, 704,
  \dodoi{10.1038/s41586-021-03394-6}

\bibitem[{{Arcodia} {et~al.}(2024{\natexlab{a}}){Arcodia}, {Liu}, {Merloni},
  {Malyali}, {Rau}, {Chakraborty}, {Goodwin}, {Buckley}, {Brink}, {Gromadzki},
  {Arzoumanian}, {Buchner}, {Kara}, {Nandra}, {Ponti}, {Salvato}, {Anderson},
  {Baldini}, {Grotova}, {Krumpe}, {Maitra}, {Miller-Jones}, \&
  {Ramos-Ceja}}]{Arcodia_2024_newQPEs}
{Arcodia}, R., {Liu}, Z., {Merloni}, A., {et~al.} 2024{\natexlab{a}}, arXiv
  e-prints, arXiv:2401.17275, \dodoi{10.48550/arXiv.2401.17275}

\bibitem[{{Arcodia} {et~al.}(2024{\natexlab{b}}){Arcodia}, {Merloni},
  {Buchner}, {Baldini}, {Ponti}, {Rau}, {Liu}, {Nandra}, \&
  {Salvato}}]{Arcodia_2024}
{Arcodia}, R., {Merloni}, A., {Buchner}, J., {et~al.} 2024{\natexlab{b}}, \aap,
  684, L14, \dodoi{10.1051/0004-6361/202348949}

\bibitem[{{Babak} {et~al.}(2017){Babak}, {Gair}, {Sesana}, {Barausse},
  {Sopuerta}, {Berry}, {Berti}, {Amaro-Seoane}, {Petiteau}, \&
  {Klein}}]{Babak2017}
{Babak}, S., {Gair}, J., {Sesana}, A., {et~al.} 2017, \prd, 95, 103012,
  \dodoi{10.1103/PhysRevD.95.103012}

\bibitem[{{Bahcall} \& {Wolf}(1976)}]{Bahcall_1976}
{Bahcall}, J.~N., \& {Wolf}, R.~A. 1976, \apj, 209, 214, \dodoi{10.1086/154711}

\bibitem[{{Bahcall} \& {Wolf}(1977)}]{Bahcall_Wolf_1977}
---. 1977, \apj, 216, 883, \dodoi{10.1086/155534}

\bibitem[{Balberg \& Yassur(2023)}]{Balberg_2023}
Balberg, S., \& Yassur, G. 2023, The Astrophysical Journal, 952, 149,
  \dodoi{10.3847/1538-4357/acdd73}

\bibitem[{{Bar-Or} \& {Alexander}(2016)}]{Bar-Or_2016}
{Bar-Or}, B., \& {Alexander}, T. 2016, \apj, 820, 129,
  \dodoi{10.3847/0004-637X/820/2/129}

\bibitem[{{Barack} \& {Cutler}(2007)}]{Barack_2007}
{Barack}, L., \& {Cutler}, C. 2007, \prd, 75, 042003,
  \dodoi{10.1103/PhysRevD.75.042003}

\bibitem[{{Binney} \& {Tremaine}(1987)}]{Binney_Tremaine}
{Binney}, J., \& {Tremaine}, S. 1987, {Galactic dynamics}

\bibitem[{{Bode} \& {Wegg}(2014)}]{Bode_2014}
{Bode}, J.~N., \& {Wegg}, C. 2014, \mnras, 438, 573,
  \dodoi{10.1093/mnras/stt2227}

\bibitem[{{Bortolas} \& {Mapelli}(2019)}]{Bortolas2019}
{Bortolas}, E., \& {Mapelli}, M. 2019, \mnras, 485, 2125,
  \dodoi{10.1093/mnras/stz440}

\bibitem[{{Broggi} {et~al.}(2022){Broggi}, {Bortolas}, {Bonetti}, {Sesana}, \&
  {Dotti}}]{Broggi_2022}
{Broggi}, L., {Bortolas}, E., {Bonetti}, M., {Sesana}, A., \& {Dotti}, M. 2022,
  \mnras, 514, 3270, \dodoi{10.1093/mnras/stac1453}

\bibitem[{{Cohn} \& {Kulsrud}(1978)}]{Cohn_Kulsrud_1978}
{Cohn}, H., \& {Kulsrud}, R.~M. 1978, \apj, 226, 1087, \dodoi{10.1086/156685}

\bibitem[{{eLISA Consortium} {et~al.}(2013){eLISA Consortium}, {Amaro Seoane},
  {Aoudia}, {Audley}, {Auger}, {Babak}, {Baker}, {Barausse}, {Barke}, {Bassan},
  {Beckmann}, {Benacquista}, {Bender}, {Berti}, {Bin{\'e}truy}, {Bogenstahl},
  {Bonvin}, {Bortoluzzi}, {Brause}, {Brossard}, {Buchman}, {Bykov}, {Camp},
  {Caprini}, {Cavalleri}, {Cerdonio}, {Ciani}, {Colpi}, {Congedo}, {Conklin},
  {Cornish}, {Danzmann}, {de Vine}, {DeBra}, {Dewi Freitag}, {Di Fiore}, {Diaz
  Aguilo}, {Diepholz}, {Dolesi}, {Dotti}, {Fern{\'a}ndez Barranco},
  {Ferraioli}, {Ferroni}, {Finetti}, {Fitzsimons}, {Gair}, {Galeazzi},
  {Garcia}, {Gerberding}, {Gesa}, {Giardini}, {Gibert}, {Grimani}, {Groot},
  {Guzman Cervantes}, {Haiman}, {Halloin}, {Heinzel}, {Hewitson}, {Hogan},
  {Holz}, {Hornstrup}, {Hoyland}, {Hoyle}, {Hueller}, {Hughes}, {Jetzer},
  {Kalogera}, {Karnesis}, {Kilic}, {Killow}, {Klipstein}, {Kochkina},
  {Korsakova}, {Krolak}, {Larson}, {Lieser}, {Littenberg}, {Livas}, {Lloro},
  {Mance}, {Madau}, {Maghami}, {Mahrdt}, {Marsh}, {Mateos}, {Mayer},
  {McClelland}, {McKenzie}, {McWilliams}, {Merkowitz}, {Miller}, {Mitryk},
  {Moerschell}, {Mohanty}, {Monsky}, {Mueller}, {M{\"u}ller}, {Nelemans},
  {Nicolodi}, {Nissanke}, {Nofrarias}, {Numata}, {Ohme}, {Otto},
  {Perreur-Lloyd}, {Petiteau}, {Phinney}, {Plagnol}, {Pollack}, {Porter},
  {Prat}, {Preston}, {Prince}, {Reiche}, {Richstone}, {Robertson}, {Rossi},
  {Rosswog}, {Rubbo}, {Ruiter}, {Sanjuan}, {Sathyaprakash}, {Schlamminger},
  {Schutz}, {Sch{\"u}tze}, {Sesana}, {Shaddock}, {Shah}, {Sheard}, {Sopuerta},
  {Spector}, {Spero}, {Stanga}, {Stebbins}, {Stede}, {Steier}, {Sumner}, {Sun},
  {Sutton}, {Tanaka}, {Tanner}, {Thorpe}, {Tr{\"o}bs}, {Tinto}, {Tu},
  {Vallisneri}, {Vetrugno}, {Vitale}, {Volonteri}, {Wand}, {Wang}, {Wanner},
  {Ward}, {Ware}, {Wass}, {Weber}, {Yu}, {Yunes}, \&
  {Zweifel}}]{eLISA_Consort2013}
{eLISA Consortium}, {Amaro Seoane}, P., {Aoudia}, S., {et~al.} 2013, arXiv
  e-prints, arXiv:1305.5720.
\newblock \doarXiv{1305.5720}

\bibitem[{Fragione \& Sari(2018)}]{Fragione_Sari_2018}
Fragione, G., \& Sari, R. 2018, The Astrophysical Journal, 852, 51,
  \dodoi{10.3847/1538-4357/aaa0d7}

\bibitem[{{Franchini} {et~al.}(2023){Franchini}, {Bonetti}, {Lupi}, {Miniutti},
  {Bortolas}, {Giustini}, {Dotti}, {Sesana}, {Arcodia}, \&
  {Ryu}}]{Franchini_2023}
{Franchini}, A., {Bonetti}, M., {Lupi}, A., {et~al.} 2023, \aap, 675, A100,
  \dodoi{10.1051/0004-6361/202346565}

\bibitem[{{Freitag}(2001)}]{Freitag_2001}
{Freitag}, M. 2001, Classical and Quantum Gravity, 18, 4033,
  \dodoi{10.1088/0264-9381/18/19/309}

\bibitem[{Gair {et~al.}(2017)Gair, Babak, Sesana, Amaro-Seoane, Barausse,
  Berry, Berti, \& Sopuerta}]{Gair_2017}
Gair, J.~R., Babak, S., Sesana, A., {et~al.} 2017, Journal of Physics:
  Conference Series, 840, 012021, \dodoi{10.1088/1742-6596/840/1/012021}

\bibitem[{{Gair} {et~al.}(2011){Gair}, {Sesana}, {Berti}, \&
  {Volonteri}}]{Gair_Sesana_2011}
{Gair}, J.~R., {Sesana}, A., {Berti}, E., \& {Volonteri}, M. 2011, Classical
  and Quantum Gravity, 28, 094018, \dodoi{10.1088/0264-9381/28/9/094018}

\bibitem[{{Gair} {et~al.}(2010){Gair}, {Tang}, \& {Volonteri}}]{Gair_2010}
{Gair}, J.~R., {Tang}, C., \& {Volonteri}, M. 2010, \prd, 81, 104014,
  \dodoi{10.1103/PhysRevD.81.104014}

\bibitem[{{Gair} {et~al.}(2013){Gair}, {Vallisneri}, {Larson}, \&
  {Baker}}]{Gair_2013}
{Gair}, J.~R., {Vallisneri}, M., {Larson}, S.~L., \& {Baker}, J.~G. 2013,
  Living Reviews in Relativity, 16, 7, \dodoi{10.12942/lrr-2013-7}

\bibitem[{{Gezari} {et~al.}(2009){Gezari}, {Heckman}, {Cenko}, {Eracleous},
  {Forster}, {Gon{\c{c}}alves}, {Martin}, {Morrissey}, {Neff}, {Seibert},
  {Schiminovich}, \& {Wyder}}]{Gezari_2009}
{Gezari}, S., {Heckman}, T., {Cenko}, S.~B., {et~al.} 2009, \apj, 698, 1367,
  \dodoi{10.1088/0004-637X/698/2/1367}

\bibitem[{{Ghez} {et~al.}(2008){Ghez}, {Salim}, {Weinberg}, {Lu}, {Do}, {Dunn},
  {Matthews}, {Morris}, {Yelda}, {Becklin}, {Kremenek}, {Milosavljevic}, \&
  {Naiman}}]{Ghez_2008}
{Ghez}, A.~M., {Salim}, S., {Weinberg}, N.~N., {et~al.} 2008, \apj, 689, 1044,
  \dodoi{10.1086/592738}

\bibitem[{{Gillessen} {et~al.}(2009){Gillessen}, {Eisenhauer}, {Trippe},
  {Alexander}, {Genzel}, {Martins}, \& {Ott}}]{Gillessen_2009}
{Gillessen}, S., {Eisenhauer}, F., {Trippe}, S., {et~al.} 2009, \apj, 692,
  1075, \dodoi{10.1088/0004-637X/692/2/1075}

\bibitem[{{Giustini} {et~al.}(2020){Giustini}, {Miniutti}, \&
  {Saxton}}]{Giustini_2020}
{Giustini}, M., {Miniutti}, G., \& {Saxton}, R.~D. 2020, \aap, 636, L2,
  \dodoi{10.1051/0004-6361/202037610}

\bibitem[{{Glampedakis} \& {Babak}(2006)}]{Glampedakis_2006}
{Glampedakis}, K., \& {Babak}, S. 2006, Classical and Quantum Gravity, 23,
  4167, \dodoi{10.1088/0264-9381/23/12/013}

\bibitem[{{Hills}(1988)}]{Hills1988}
{Hills}, J.~G. 1988, \nat, 331, 687, \dodoi{10.1038/331687a0}

\bibitem[{{Hils} \& {Bender}(1995)}]{Hils_1995}
{Hils}, D., \& {Bender}, P.~L. 1995, \apjl, 445, L7, \dodoi{10.1086/187876}

\bibitem[{{Hopman} \& {Alexander}(2005)}]{Hopman_Alexander_2005}
{Hopman}, C., \& {Alexander}, T. 2005, \apj, 629, 362, \dodoi{10.1086/431475}

\bibitem[{{Hopman} \& {Alexander}(2006)}]{Hopman_2006}
---. 2006, \apjl, 645, L133, \dodoi{10.1086/506273}

\bibitem[{{Kaur} {et~al.}(2018){Kaur}, {Kazandjian}, {Sridhar}, \&
  {Touma}}]{Kaur_2018}
{Kaur}, K., {Kazandjian}, M.~V., {Sridhar}, S., \& {Touma}, J.~R. 2018, \mnras,
  476, 4104, \dodoi{10.1093/mnras/sty403}

\bibitem[{{Kaur} \& {Perets}(in prep.)}]{Kaur_Perets_24}
{Kaur}, K., \& {Perets}, H. in prep.

\bibitem[{{Kaur} {et~al.}(2023){Kaur}, {Stone}, \& {Gilbaum}}]{Kaur_Stone_2023}
{Kaur}, K., {Stone}, N.~C., \& {Gilbaum}, S. 2023, \mnras, 524, 1269,
  \dodoi{10.1093/mnras/stad1894}

\bibitem[{{Keshet} {et~al.}(2009){Keshet}, {Hopman}, \&
  {Alexander}}]{Keshet_2009}
{Keshet}, U., {Hopman}, C., \& {Alexander}, T. 2009, \apjl, 698, L64,
  \dodoi{10.1088/0004-637X/698/1/L64}

\bibitem[{{Kormendy} \& {Ho}(2013)}]{Kormendy_Ho_2013}
{Kormendy}, J., \& {Ho}, L.~C. 2013, \araa, 51, 511,
  \dodoi{10.1146/annurev-astro-082708-101811}

\bibitem[{{Kulkarni} {et~al.}(2021){Kulkarni}, {Harrison}, {Grefenstette},
  {Earnshaw}, {Andreoni}, {Berg}, {Bloom}, {Cenko}, {Chornock}, {Christiansen},
  {Coughlin}, {Wuollet Criswell}, {Darvish}, {Das}, {De}, {Dessart}, {Dixon},
  {Dorsman}, {El-Badry}, {Evans}, {Ford}, {Fremling}, {Gansicke}, {Gezari},
  {Goetberg}, {Green}, {Graham}, {Heida}, {Ho}, {Jaodand}, {Johns-Krull},
  {Kasliwal}, {Lazzarini}, {Lu}, {Margutti}, {Martin}, {Masters}, {McKernan},
  {Naze}, {Nissanke}, {Parazin}, {Perley}, {Phinney}, {Piro}, {Raaijmakers},
  {Rauw}, {Rodriguez}, {Sana}, {Senchyna}, {Singer}, {Spake}, {Stassun},
  {Stern}, {Teplitz}, {Weisz}, \& {Yao}}]{Kulkarni_2021}
{Kulkarni}, S.~R., {Harrison}, F.~A., {Grefenstette}, B.~W., {et~al.} 2021,
  arXiv e-prints, arXiv:2111.15608, \dodoi{10.48550/arXiv.2111.15608}

\bibitem[{{Levin}(2007)}]{Levin_2007}
{Levin}, Y. 2007, \mnras, 374, 515, \dodoi{10.1111/j.1365-2966.2006.11155.x}

\bibitem[{{Lightman} \& {Shapiro}(1977)}]{Lightman_Shapiro1977}
{Lightman}, A.~P., \& {Shapiro}, S.~L. 1977, \apj, 211, 244,
  \dodoi{10.1086/154925}

\bibitem[{{Linial} \& {Metzger}(2023)}]{Linial_Metzger_2023}
{Linial}, I., \& {Metzger}, B.~D. 2023, \apj, 957, 34,
  \dodoi{10.3847/1538-4357/acf65b}

\bibitem[{{Linial} \& {Metzger}(2024)}]{Linial_2024_uv}
---. 2024, \apjl, 963, L1, \dodoi{10.3847/2041-8213/ad2464}

\bibitem[{{Linial} \& {Sari}(2017)}]{Linial_Sari_2017}
{Linial}, I., \& {Sari}, R. 2017, \mnras, 469, 2441,
  \dodoi{10.1093/mnras/stx1041}

\bibitem[{{Linial} \& {Sari}(2022)}]{Linial_Sari_22}
---. 2022.
\newblock \doarXiv{2206.14817}

\bibitem[{{Linial} \& {Sari}(2023)}]{Linial_Sari_2023}
---. 2023, \apj, 945, 86, \dodoi{10.3847/1538-4357/acbd3d}

\bibitem[{{Madigan} {et~al.}(2009){Madigan}, {Levin}, \&
  {Hopman}}]{Madigan_2009}
{Madigan}, A.-M., {Levin}, Y., \& {Hopman}, C. 2009, \apjl, 697, L44,
  \dodoi{10.1088/0004-637X/697/1/L44}

\bibitem[{{Mazzolari} {et~al.}(2022){Mazzolari}, {Bonetti}, {Sesana},
  {Colombo}, {Dotti}, {Lodato}, \& {Izquierdo-Villalba}}]{Mazzolari_2022}
{Mazzolari}, G., {Bonetti}, M., {Sesana}, A., {et~al.} 2022, \mnras, 516, 1959,
  \dodoi{10.1093/mnras/stac2255}

\bibitem[{{Mei} {et~al.}(2021){Mei}, {Bai}, {Bao}, {Barausse}, {Cai}, {Canuto},
  {Cao}, {Chen}, {Chen}, {Ding}, {Duan}, {Fan}, {Feng}, {Fu}, {Gao}, {Gao},
  {Gong}, {Gou}, {Gu}, {Gu}, {He}, {Hendry}, {Hong}, {Hu}, {Hu}, {Hu}, {Huang},
  {Huang}, {Jiang}, {Jiang}, {Jiang}, {Jiang}, {Jin}, {Korol}, {Li}, {Li},
  {Li}, {Li}, {Li}, {Li}, {Li}, {Li}, {Li}, {Liang}, {Liang}, {Liao}, {Liu},
  {Liu}, {Liu}, {Liu}, {Liu}, {Liu}, {Liu}, {Lu}, {Lu}, {Lu}, {Luo}, {Luo},
  {Milyukov}, {Ming}, {Pi}, {Qin}, {Qu}, {Sesana}, {Shao}, {Shi}, {Su}, {Tan},
  {Tan}, {Tan}, {Tu}, {Wang}, {Wang}, {Wang}, {Wang}, {Wang}, {Wang}, {Wang},
  {Wang}, {Wang}, {Wang}, {Wang}, {Wei}, {Wu}, {Xiao}, {Xu}, {Xue}, {Yang},
  {Yang}, {Yang}, {Yang}, {Ye}, {Yeh}, {Yu}, {Zhai}, {Zhang}, {Zhang}, {Zhang},
  {Zhang}, {Zhang}, {Zhang}, {Zhang}, {Zhou}, {Zhou}, {Zhou}, {Zhu}, {Zi}, \&
  {Luo}}]{Mei_2021}
{Mei}, J., {Bai}, Y.-Z., {Bao}, J., {et~al.} 2021, Progress of Theoretical and
  Experimental Physics, 2021, 05A107, \dodoi{10.1093/ptep/ptaa114}

\bibitem[{{Merritt}(2013)}]{Merritt_2013}
{Merritt}, D. 2013, {Dynamics and Evolution of Galactic Nuclei}

\bibitem[{{Merritt}(2015)}]{Merritt_2015}
---. 2015, \apj, 814, 57, \dodoi{10.1088/0004-637X/814/1/57}

\bibitem[{{Merritt} {et~al.}(2011){Merritt}, {Alexander}, {Mikkola}, \&
  {Will}}]{Merritt_2011}
{Merritt}, D., {Alexander}, T., {Mikkola}, S., \& {Will}, C.~M. 2011, \prd, 84,
  044024, \dodoi{10.1103/PhysRevD.84.044024}

\bibitem[{{Metzger} {et~al.}(2022){Metzger}, {Stone}, \&
  {Gilbaum}}]{Metzger_2022}
{Metzger}, B.~D., {Stone}, N.~C., \& {Gilbaum}, S. 2022, \apj, 926, 101,
  \dodoi{10.3847/1538-4357/ac3ee1}

\bibitem[{{Miller} {et~al.}(2005){Miller}, {Freitag}, {Hamilton}, \&
  {Lauburg}}]{Miller_2005}
{Miller}, M.~C., {Freitag}, M., {Hamilton}, D.~P., \& {Lauburg}, V.~M. 2005,
  \apjl, 631, L117, \dodoi{10.1086/497335}

\bibitem[{{Miniutti} {et~al.}(2019){Miniutti}, {Saxton}, {Giustini},
  {Alexander}, {Fender}, {Heywood}, {Monageng}, {Coriat}, {Tzioumis}, {Read},
  {Knigge}, {Gandhi}, {Pretorius}, \& {Ag{\'\i}s-Gonz{\'a}lez}}]{Miniutti_2019}
{Miniutti}, G., {Saxton}, R.~D., {Giustini}, M., {et~al.} 2019, \nat, 573, 381,
  \dodoi{10.1038/s41586-019-1556-x}

\bibitem[{{Nandra} {et~al.}(2013){Nandra}, {Barret}, {Barcons}, {Fabian}, {den
  Herder}, {Piro}, {Watson}, {Adami}, {Aird}, {Afonso}, {Alexander},
  {Argiroffi}, {Amati}, {Arnaud}, {Atteia}, {Audard}, {Badenes}, {Ballet},
  {Ballo}, {Bamba}, {Bhardwaj}, {Stefano Battistelli}, {Becker}, {De Becker},
  {Behar}, {Bianchi}, {Biffi}, {B{\^\i}rzan}, {Bocchino}, {Bogdanov}, {Boirin},
  {Boller}, {Borgani}, {Borm}, {Bouch{\'e}}, {Bourdin}, {Bower}, {Braito},
  {Branchini}, {Branduardi-Raymont}, {Bregman}, {Brenneman}, {Brightman},
  {Br{\"u}ggen}, {Buchner}, {Bulbul}, {Brusa}, {Bursa}, {Caccianiga},
  {Cackett}, {Campana}, {Cappelluti}, {Cappi}, {Carrera}, {Ceballos},
  {Christensen}, {Chu}, {Churazov}, {Clerc}, {Corbel}, {Corral}, {Comastri},
  {Costantini}, {Croston}, {Dadina}, {D'Ai}, {Decourchelle}, {Della Ceca},
  {Dennerl}, {Dolag}, {Done}, {Dovciak}, {Drake}, {Eckert}, {Edge}, {Ettori},
  {Ezoe}, {Feigelson}, {Fender}, {Feruglio}, {Finoguenov}, {Fiore}, {Galeazzi},
  {Gallagher}, {Gandhi}, {Gaspari}, {Gastaldello}, {Georgakakis},
  {Georgantopoulos}, {Gilfanov}, {Gitti}, {Gladstone}, {Goosmann}, {Gosset},
  {Grosso}, {Guedel}, {Guerrero}, {Haberl}, {Hardcastle}, {Heinz}, {Alonso
  Herrero}, {Herv{\'e}}, {Holmstrom}, {Iwasawa}, {Jonker}, {Kaastra}, {Kara},
  {Karas}, {Kastner}, {King}, {Kosenko}, {Koutroumpa}, {Kraft}, {Kreykenbohm},
  {Lallement}, {Lanzuisi}, {Lee}, {Lemoine-Goumard}, {Lobban}, {Lodato},
  {Lovisari}, {Lotti}, {McCharthy}, {McNamara}, {Maggio}, {Maiolino}, {De
  Marco}, {de Martino}, {Mateos}, {Matt}, {Maughan}, {Mazzotta}, {Mendez},
  {Merloni}, {Micela}, {Miceli}, {Mignani}, {Miller}, {Miniutti}, {Molendi},
  {Montez}, {Moretti}, {Motch}, {Naz{\'e}}, {Nevalainen}, {Nicastro}, {Nulsen},
  {Ohashi}, {O'Brien}, {Osborne}, {Oskinova}, {Pacaud}, {Paerels}, {Page},
  {Papadakis}, {Pareschi}, {Petre}, {Petrucci}, {Piconcelli}, {Pillitteri},
  {Pinto}, {de Plaa}, {Pointecouteau}, {Ponman}, {Ponti}, {Porquet}, {Pounds},
  {Pratt}, {Predehl}, {Proga}, {Psaltis}, {Rafferty}, {Ramos-Ceja}, {Ranalli},
  {Rasia}, {Rau}, {Rauw}, {Rea}, {Read}, {Reeves}, {Reiprich}, {Renaud},
  {Reynolds}, {Risaliti}, {Rodriguez}, {Rodriguez Hidalgo}, {Roncarelli},
  {Rosario}, {Rossetti}, {Rozanska}, {Rovilos}, {Salvaterra}, {Salvato}, {Di
  Salvo}, {Sanders}, {Sanz-Forcada}, {Schawinski}, {Schaye}, {Schwope},
  {Sciortino}, {Severgnini}, {Shankar}, {Sijacki}, {Sim}, {Schmid}, {Smith},
  {Steiner}, {Stelzer}, {Stewart}, {Strohmayer}, {Str{\"u}der}, {Sun}, {Takei},
  {Tatischeff}, {Tiengo}, {Tombesi}, {Trinchieri}, {Tsuru}, {Ud-Doula},
  {Ursino}, {Valencic}, {Vanzella}, {Vaughan}, {Vignali}, {Vink}, {Vito},
  {Volonteri}, {Wang}, {Webb}, {Willingale}, {Wilms}, {Wise}, {Worrall},
  {Young}, {Zampieri}, {In't Zand}, {Zane}, {Zezas}, {Zhang}, \&
  {Zhuravleva}}]{Nandra_2013}
{Nandra}, K., {Barret}, D., {Barcons}, X., {et~al.} 2013, arXiv e-prints,
  arXiv:1306.2307, \dodoi{10.48550/arXiv.1306.2307}

\bibitem[{{Naoz} {et~al.}(2022){Naoz}, {Rose}, {Michaely}, {Melchor},
  {Ramirez-Ruiz}, {Mockler}, \& {Schnittman}}]{Naoz_2022}
{Naoz}, S., {Rose}, S.~C., {Michaely}, E., {et~al.} 2022, \apjl, 927, L18,
  \dodoi{10.3847/2041-8213/ac574b}

\bibitem[{{Neumayer} {et~al.}(2020){Neumayer}, {Seth}, \&
  {B{\"o}ker}}]{Neumayer_2020}
{Neumayer}, N., {Seth}, A., \& {B{\"o}ker}, T. 2020, \aapr, 28, 4,
  \dodoi{10.1007/s00159-020-00125-0}

\bibitem[{{Pan} {et~al.}(2022){Pan}, {Li}, {Cao}, {Miniutti}, \&
  {Gu}}]{Pan_2022_viscosity_model}
{Pan}, X., {Li}, S.-L., {Cao}, X., {Miniutti}, G., \& {Gu}, M. 2022, \apjl,
  928, L18, \dodoi{10.3847/2041-8213/ac5faf}

\bibitem[{{Pan} {et~al.}(2021){Pan}, {Lyu}, \& {Yang}}]{Pan2021}
{Pan}, Z., {Lyu}, Z., \& {Yang}, H. 2021, \prd, 104, 063007,
  \dodoi{10.1103/PhysRevD.104.063007}

\bibitem[{{Perets} {et~al.}(2007){Perets}, {Hopman}, \&
  {Alexander}}]{Perets2007}
{Perets}, H.~B., {Hopman}, C., \& {Alexander}, T. 2007, \apj, 656, 709,
  \dodoi{10.1086/510377}

\bibitem[{{Peters}(1964)}]{Peters_1964}
{Peters}, P.~C. 1964, Physical Review, 136, 1224,
  \dodoi{10.1103/PhysRev.136.B1224}

\bibitem[{{Predehl} {et~al.}(2021){Predehl}, {Andritschke}, {Arefiev},
  {Babyshkin}, {Batanov}, {Becker}, {B{\"o}hringer}, {Bogomolov}, {Boller},
  {Borm}, {Bornemann}, {Br{\"a}uninger}, {Br{\"u}ggen}, {Brunner}, {Brusa},
  {Bulbul}, {Buntov}, {Burwitz}, {Burkert}, {Clerc}, {Churazov}, {Coutinho},
  {Dauser}, {Dennerl}, {Doroshenko}, {Eder}, {Emberger}, {Eraerds},
  {Finoguenov}, {Freyberg}, {Friedrich}, {Friedrich}, {F{\"u}rmetz},
  {Georgakakis}, {Gilfanov}, {Granato}, {Grossberger}, {Gueguen}, {Gureev},
  {Haberl}, {H{\"a}lker}, {Hartner}, {Hasinger}, {Huber}, {Ji}, {Kienlin},
  {Kink}, {Korotkov}, {Kreykenbohm}, {Lamer}, {Lomakin}, {Lapshov}, {Liu},
  {Maitra}, {Meidinger}, {Menz}, {Merloni}, {Mernik}, {Mican}, {Mohr},
  {M{\"u}ller}, {Nandra}, {Nazarov}, {Pacaud}, {Pavlinsky}, {Perinati},
  {Pfeffermann}, {Pietschner}, {Ramos-Ceja}, {Rau}, {Reiffers}, {Reiprich},
  {Robrade}, {Salvato}, {Sanders}, {Santangelo}, {Sasaki}, {Scheuerle},
  {Schmid}, {Schmitt}, {Schwope}, {Shirshakov}, {Steinmetz}, {Stewart},
  {Str{\"u}der}, {Sunyaev}, {Tenzer}, {Tiedemann}, {Tr{\"u}mper}, {Voron},
  {Weber}, {Wilms}, \& {Yaroshenko}}]{Predehl_2021}
{Predehl}, P., {Andritschke}, R., {Arefiev}, V., {et~al.} 2021, \aap, 647, A1,
  \dodoi{10.1051/0004-6361/202039313}

\bibitem[{{Qunbar} \& {Stone}(2023)}]{Qunbar_Stone_2023}
{Qunbar}, I., \& {Stone}, N.~C. 2023, arXiv e-prints, arXiv:2304.13062,
  \dodoi{10.48550/arXiv.2304.13062}

\bibitem[{{Raj} {et~al.}(2021){Raj}, {Nixon}, \& {Do{\u{g}}an}}]{Raj2021}
{Raj}, A., {Nixon}, C.~J., \& {Do{\u{g}}an}, S. 2021, \apj, 909, 81,
  \dodoi{10.3847/1538-4357/abdc24}

\bibitem[{{Rauch} \& {Tremaine}(1996)}]{Rauch_Tremaine_1996}
{Rauch}, K.~P., \& {Tremaine}, S. 1996, \na, 1, 149,
  \dodoi{10.1016/S1384-1076(96)00012-7}

\bibitem[{{Raveh} \& {Perets}(2021)}]{Raveh_2021}
{Raveh}, Y., \& {Perets}, H.~B. 2021, \mnras, 501, 5012,
  \dodoi{10.1093/mnras/staa4001}

\bibitem[{{Rom} {et~al.}(in prep.){Rom}, {Linial}, {Kaur}, \& {Sari}}]{Rom_24}
{Rom}, B., {Linial}, I., {Kaur}, K., \& {Sari}, R. in prep.

\bibitem[{Rose {et~al.}(2023)Rose, Naoz, Sari, \& Linial}]{Rose_2023}
Rose, S.~C., Naoz, S., Sari, R., \& Linial, I. 2023, The Astrophysical Journal,
  955, 30, \dodoi{10.3847/1538-4357/acee75}

\bibitem[{{Ryan}(1995)}]{Ryan_1995}
{Ryan}, F.~D. 1995, \prd, 52, 5707, \dodoi{10.1103/PhysRevD.52.5707}

\bibitem[{{Ryan}(1997)}]{Ryan1997}
---. 1997, \prd, 56, 1845, \dodoi{10.1103/PhysRevD.56.1845}

\bibitem[{{Sagiv} {et~al.}(2014){Sagiv}, {Gal-Yam}, {Ofek}, {Waxman},
  {Aharonson}, {Kulkarni}, {Nakar}, {Maoz}, {Trakhtenbrot}, {Phinney}, {Topaz},
  {Beichman}, {Murthy}, \& {Worden}}]{Sagiv_2014}
{Sagiv}, I., {Gal-Yam}, A., {Ofek}, E.~O., {et~al.} 2014, \aj, 147, 79,
  \dodoi{10.1088/0004-6256/147/4/79}

\bibitem[{{Sari} \& {Fragione}(2019)}]{Sari_Fragione_2019}
{Sari}, R., \& {Fragione}, G. 2019, \apj, 885, 24,
  \dodoi{10.3847/1538-4357/ab43df}

\bibitem[{{Sigurdsson} \& {Rees}(1997)}]{Sigurdsson_1997}
{Sigurdsson}, S., \& {Rees}, M.~J. 1997, \mnras, 284, 318,
  \dodoi{10.1093/mnras/284.2.318}

\bibitem[{{Tagawa} \& {Haiman}(2023)}]{Tagawa_2023}
{Tagawa}, H., \& {Haiman}, Z. 2023, \mnras, 526, 69,
  \dodoi{10.1093/mnras/stad2616}

\bibitem[{{Teboul} {et~al.}(2024){Teboul}, {Stone}, \&
  {Ostriker}}]{Teboul_2024}
{Teboul}, O., {Stone}, N.~C., \& {Ostriker}, J.~P. 2024, \mnras, 527, 3094,
  \dodoi{10.1093/mnras/stad3301}

\bibitem[{{Xian} {et~al.}(2021){Xian}, {Zhang}, {Dou}, {He}, \&
  {Shu}}]{Xian_2021}
{Xian}, J., {Zhang}, F., {Dou}, L., {He}, J., \& {Shu}, X. 2021, \apjl, 921,
  L32, \dodoi{10.3847/2041-8213/ac31aa}

\bibitem[{{Yu} \& {Tremaine}(2003)}]{Yu_Tremaine_2003}
{Yu}, Q., \& {Tremaine}, S. 2003, \apj, 599, 1129, \dodoi{10.1086/379546}

\bibitem[{{Yuan} {et~al.}(2022){Yuan}, {Zhang}, {Chen}, \& {Ling}}]{Yuan_2022}
{Yuan}, W., {Zhang}, C., {Chen}, Y., \& {Ling}, Z. 2022, in Handbook of X-ray
  and Gamma-ray Astrophysics, 86, \dodoi{10.1007/978-981-16-4544-0_151-1}

\end{thebibliography}
\bibliographystyle{aasjournal}

\appendix

\section{Analytical Solutions} 
\label{app_ana_sol1}

The second order ODE~(\ref{ode_in_g}) in $g(x)$ can be analytically integrated for $\beta = 0,5/2$. These cases correspond to $\gamma = 3/2 , 11/4$ for the choice of BW profile of field stars with $\gamf = 7/4$. Here we provide these analytical solutions.

\medskip

\noindent
{{\bf Case A.} $\beta = 5/2$ (equivalently $\gamma = 11/4$ for $\gamf = 7/4$) } 

\smallskip 

For $\beta = 5/2$, it is straight-forward to integrate ODE~(\ref{ode_in_g}) over $x$ twice, to get the following general solution:
\beq
 g(x)  = - \frac{4 (3 - \gamf)}{25} B  \exp{(y)}   \Ei(-y) + C \exp{(y)} \qquad 
 \mbox{where  } \quad y = \frac{4 (3 -\gamf)}{ 25 x^{5/2}} \, 
\label{g_ana_gen1}
\eeq 
and $B$ and $C$ are constants of integration. $\Ei$ is the standard exponential integral function with the definition (and limiting functional forms to be used later) given  as:
\beq 
\Ei(-y) = - \int_{y}^{\infty} \rmd t \frac{\exp{(-t)}}{t} = \begin{cases}
    \gamE + \log{|y|} \; \, \quad, \quad y \rightarrow 0  \\
    \displaystyle{- \frac{\exp{(-y)}}{y} }  \quad , \quad y \rightarrow \infty  \\ 
\end{cases} \, 
\label{exp_int_defs}
\eeq 
where $\gamE \simeq 0.577216$ is the Euler's constant.   

Now, we determine the constants $\{B,C \}$ that satisfy the intended forms of asymptotic solutions $g_i(x) = \log{x}+c_0$ (for $x \gg 1$) and $\ggw(x) = 5 c_1 x^{\beta}/(2 (3- \gamf)) $ (for $x \ll 1$); see equation~(\ref{g_gw_gen}). 

For $x \gg 1 $ (or $y \rightarrow 0$), the general solution reduces to the form: 
\beq 
g(x) =  \frac{2 (3 - \gamf) B}{5} \log{x} + C - \frac{4 (3 - \gamf) B}{25 }  \bigg\{ \gamE + \log{\bigg( \frac{4 (3-\gamf)}{25}  \bigg)}  \bigg\} 
\eeq 
Comparing the above limiting solution with $g_i(x)$, we have: 
\beq 
 B = \frac{5}{2 (3 - \gamf)} \qquad , \qquad 
 C = c_0 + \frac{4 (3 - \gamf) B}{25 }  \bigg\{ \gamE + \log{\bigg( \frac{4 (3-\gamf)}{25}  \bigg)}  \bigg\} 
\label{B_C_values}
\eeq 

Similarly, for $x \ll 1 $ (or $y \rightarrow \infty$), the general solution reduces to the form:
\beq 
g(x) = B x^{5/2} + C \exp{\bigg( \frac{4 (3-\gamf)}{25 x^{5/2}} \bigg)}
\eeq 
Since the second term diverges for $x \rightarrow 0$, it is not a physically interesting solution and hence $C=0$. Also, the remaining first term matches with the functional form of intended asymptotic solution $\ggw$, which gives $c_1 = 2 B (3-\gamf)/5$. Further, using equation~(\ref{B_C_values}), we have the required numerical constants: 
\beq 
 c_1 = 1 \qquad , \qquad  c_0 = - \frac{2}{5} \gamE - \frac{2}{5} \log{\bigg( \frac{4 (3 - \gamf)}{25} \bigg)} 
\label{c1_c0_gamma11/4}
\eeq 
From equation~(\ref{g_ana_gen1}), the required solution for $\beta = 5/2$ that satisfies the intended boundary conditions is given as: 
\beq 
g(x)  = - \frac{2 }{5} \exp{(y)}   \Ei(-y)  \qquad 
 \mbox{where  } \quad y = \frac{4 (3 -\gamf)}{ 25 x^{5/2}}
\eeq 

\medskip

\noindent
{{\bf Case B.} $\beta = 0$ (equivalently $\gamma = 3/2$) } 

\smallskip 

For $\beta = 0$, the ODE~(\ref{ode_in_g}) can be integrated twice in $x$, resulting in the following form of general solution: 
\beq 
g(x) = B' - \frac{2}{5} C' \, \Ei{(y)} \quad, \quad  \mbox{where  } \quad y = \frac{4 (3 -\gamf)}{ 25 x^{5/2}} 
\label{g_ana_gen2} 
\eeq 
where $B'$ and $C'$ are the integration constants. In the limit of $x \rightarrow 0 $, the second term diverges to $-\infty$ from equation~(\ref{exp_int_defs}). To avoid this unphysical behavior, $C' = 0$ and the required solution becomes $g(x) = B'= $ constant which matches the asymptotic solution $g_i(x)$ for the case of a zero diffusive flux $\scF_p$; see the discussion below  equation~(\ref{scNi}).

\section{Numerical Methods}
\label{app_num_meth}

Here we describe our approach to numerically solve the ODE of equation~(\ref{ode_g_special}) for the desired forms of asymptotic solutions. We change the dependent variable to $h(x) = g(x)/\ggw(x) = g(x)/(2 c_1 x^{2 \gamma - 3})$. Since $g \rightarrow \ggw$ as $x \rightarrow 0$, the suitable initial conditions to solve this ODE are $h(\xini) = 1$, $h'(\xini) = 0$ where $\xini \ll 1$. We integrate this ODE over a wide range of $x \in [\xini,\xfin]$ where $\xfin \gg 1$. We compare the numerical solution with the expected solution $h(x) = g_i(x)/\ggw(x) = (\log{x} + c_0)/(2 c_1 x^{2 \gamma - 3})$ in the large $x$ limit to deduce the numerical constants $\{c_1,c_0\}$. These constants are explicitly given by the following relations: 
\beq 
\begin{split} 
c_1 &= \frac{1}{2 \xfin^{2 \gamma - 2}} \bigg[ h'(\xfin) + \frac{(2 \gamma -3)}{\xfin} h(\xfin)  \bigg]^{-1} \\[1ex] 
c_0 &= 2 c_1 \xfin^{2 \gamma -3} h(\xfin) - \log{\xfin}
 \end{split} 
 \label{c1_c0_num}
\eeq 
We check the consistency of the deduced values of $\{c_1,c_0 \}$ by choosing different $\xini = 10^{-p}$ and $\xfin = 10^{p}$, with a wide range of $p = 2-7$. We find these values to be consistent with fractional differences lying within a few per cent.

\end{document}